\documentstyle[aps,multicol,epsfig,psfig]{revtex}
\newcommand{\bea}{\begin{eqnarray}}
\newcommand{\eea}{\end{eqnarray}}
\newcommand{\evec}{\vec{e}}
\newcommand{\vvec}{\vec{v}}
\begin{document}
\draft
\title{Spectra of ``Real--World'' Graphs: Beyond the Semi-Circle Law}

\author{
Ill\'{e}s J. Farkas$^{\, 1}$, 
Imre Der\'{e}nyi$^{\,  2,3}$, 
Albert-L\'{a}szl\'{o} Barab\'{a}si$^{\, 2,4}$, 
Tam\'{a}s Vicsek$^{\, 1,2}$
}
\address{
$^{1}$ Department of Biological Physics, E\"otv\"os University, Budapest,
P\'azm\'any P\'eter S\'et\'any 1A, H-1117 Hungary\\
$^{2}$ Collegium Budapest, Institute for Advanced Study,
Budapest, Szenth\'aroms\'ag utca 2, H-1014 Hungary\\
$^{3}$ Institut Curie, UMR 168, 26 rue d'Ulm, F-75248 Paris 05,
France\\
$^{4}$ Department of Physics, University of Notre Dame, 
Notre Dame, IN 46556\\
{\tt fij@elte.hu, derenyi@angel.elte.hu, alb@nd.edu, vicsek@angel.elte.hu}}
\maketitle
\thispagestyle{empty}
\begin{abstract}

Many natural and social systems develop complex networks, 
that are usually modelled as
random graphs. The eigenvalue spectrum of these
graphs provides information about their structural properties.
While the semi-circle law is known to describe the
spectral density of uncorrelated random graphs,
much less is known about the eigenvalues of real-world graphs,
describing such complex systems as the Internet,
metabolic pathways, networks of power stations,
scientific collaborations or movie actors, 
which are inherently correlated
and usually very sparse.
An important limitation in addressing the spectra of these systems is
that the numerical determination 
of the spectra for systems with more than a few
thousand nodes is prohibitively time and memory consuming. 
Making use of recent advances in algorithms for spectral characterization,
here we develop new methods 
to determine the eigenvalues of networks
comparable in size to real systems, obtaining
several surprising results on the spectra
of adjacency matrices corresponding to models of real-world graphs.
We find that when the number of links grows as the number of nodes,
the spectral density of uncorrelated random graphs does not
converge to the semi-circle law. 
Furthermore, the spectral densities of real-world graphs have specific
features depending on 
the details of the corresponding models. In particular, 
scale-free graphs develop a triangle-like 
spectral density with a power law tail,
while small-world graphs have a complex spectral density function consisting
of several sharp peaks. These and further results 
indicate that the spectra of
correlated graphs represent a practical tool for 
graph classification and can provide useful insight into 
the relevant structural properties of real networks.

\end{abstract}

\begin{multicols}{2}



\section{Introduction}
\label{s_intro}

Random graphs\cite{ErdosRenyi,Bollobas}
have long been used for modelling the
evolution and topology of systems made up of large assemblies of
{\it similar units}.
The uncorrelated random graph model
-- which assumes each pair of the graph's vertices
to be connected with equal and independent probabilities --
treats a network as an assembly of {\it equivalent units}. 
This model, introduced by the mathematicians 
Paul Erd\H os and Alfr\'ed R\'enyi\cite{ErdosRenyi},
has been much investigated in the mathematical literature\cite{Bollobas}.
However, the increasing availability of large maps of real-life 
networks has indicated that real networks are fundamentally correlated
systems, and in many respects their topology deviates from the
uncorrelated random graph model.
Consequently, the attention has shifted towards
{\it more advanced graph models}
which are designed to generate topologies 
in line with the existing empirical results 
\cite{Redner,WattsStrogatz,Watts_book,falou,AdamicHuberman,DiameterOfTheWWW,Metabolic,
EmergenceOfScaling,ScaleFree_PA,Amaral_ClassesOfSW,Barab_collab,Newman_collab}. 
Examples of real networks, that serve as a benchmark for the
current modelling efforts include
the Internet 
\cite{falou,JB,Cohen_error,Cohen_attack}, 
the World-Wide Web
\cite{DiameterOfTheWWW,Broder}, 
networks of collaborating movie actors
and those of collaborating scientists
\cite{Barab_collab,Newman_collab},
the power grid
\cite{WattsStrogatz,Watts_book}
 and the metabolic network of numerous living
organisms
\cite{Metabolic,WagnerFell}. 

These are the systems that 
we will call {\it ``real-world'' networks or graphs}. 
Several converging reasons explain the enhanced current interest in
such real graphs. First, the
{\it amount of topological data} available on such large structures
has increased dramatically during the past few years
thanks to the computerization of data collection in various fields,
from sociology to biology.
Second, the hitherto unseen {\it speed of growth} 
of some of these complex networks  
-- e.g., the Internet --
and their pervasiveness in affecting many aspects of our life has
created the need to understand the topology, 
origin and evolution of such structures.
Finally, the increased {\it computational power} 
available on almost every desktop has allowed, for the first time,
to study such systems in unprecedented detail.

\par 
The proliferation of data has lead to a flurry of activity towards
understanding the general properties of real networks. These efforts
have resulted in the introduction of two classes of models, 
commonly called
{\it small-world graphs}\cite{WattsStrogatz,Watts_book}
and the {\it scale-free networks}\cite{EmergenceOfScaling,ScaleFree_PA}.
The first aims to capture the clustering observed in real graphs, while
the second reproduces the power law degree distribution present in
many real networks.
However, until now, most analyses of 
these models and data sets 
have been confined to real-space characteristics,
that capture their static structural properties:
e.g., degree sequences, shortest connecting 
paths and clustering coefficients. 
In contrast, there is extensive literature demonstrating that the
properties of graphs and the associated adjacency matrices are well
characterized by spectral methods, that provide global
measures of the network properties\cite{Mehta_book,Vulpiani}.
In this paper, we offer a 
detailed analysis of the most studied network models
using algebraic tools intrinsic to large random graphs.

\par
The paper is organized as follows. 
Section\,\ref{s_model} introduces the 
main random graph models
used for the topological description of large assemblies of
connected units.
Section\,\ref{s_tools} lists the 
-- analytical and numerical -- 
{\it tools} which we used and developed 
to convert the 
topological features of graphs into algebraic invariants.  
Section\,\ref{s_res} contains our {\it results}
concerning the spectra
and special eigenvalues of the three main types of random graph
models: 
sparse uncorrelated random graphs in Section\,\ref{ss_ER.sparse}; 
small-world graphs in Section\,\ref{ss_SW}; and
scale-free networks in Section\,\ref{ss_SF}.  
Section\,\ref{ss_test}
gives simple algorithms for {\it testing the graph's structure},
and Section\,\ref{ss_var} investigates the variance of structure
within single random graph models.

\section{Models of random graphs}
\label{s_model}

\subsection{The uncorrelated random graph model and the semi-circle law}
\label{ss_uncorr}

\subsubsection{Definitions}
\label{sss_def}

Throughout this paper, 
we will use the term {\it ``graph''} for a
set of points (vertices) connected by undirected lines (edges);
no multiple edges and 
no loops connecting a vertex to itself are allowed.
We will call two vertices of the graph {\it ``neighbors''}, 
if they are connected by an edge.
Based on Ref.\,\cite{ErdosRenyi},
we shall use the term {\it ``uncorrelated random graph''} for a
graph, if (i) the probability for any pair of the graph's vertices 
being connected is the same, $p$;
(ii) these probabilities are independent variables.

\par
Any graph $G$ can be 
represented by its {\it adjacency matrix}, $A(G)$, 
which is a real symmetric matrix:
$A_{ij}=A_{ji}=1$, if vertices $i$ and $j$ are connected, 
or $0$,	if these two vertices are not connected.
The main algebraic tool which we will use for the 
analysis of graphs will be the 
spectrum -- i.e., the set of eigenvalues -- 
of the graph's adjacency matrix. 
The spectrum of the graph's adjacency matrix is  also called
the {\it spectrum of the graph}.

\subsubsection{Applying the semi-circle law for the spectrum of the
uncorrelated random graph}
\label{sss_appl}

A general form of the {\it semi-circle law} 
for real symmetric matrices is the 
following\cite{Mehta_book,Wigner,HiaiPetz}.
If $A$ is a real symmetric $N\times N$
uncorrelated random matrix, 
$\langle A_{ij}\rangle\equiv 0$ and
$\langle A_{ij}^2\rangle =\sigma^2$ 
for every $i\not= j$,
and with increasing $N$ each moment of each 
$|A_{ij}|$ remains finite,
then in the $N\to\infty$ limit the spectral density
-- i.e., the density of eigenvalues --
of $A/\sqrt{N}$ converges to the 
{\it semi-circular distribution}:

\bea
\rho(\lambda) = \
\cases {
(2\pi\sigma^2)^{-1} \sqrt{4\sigma^2-\lambda^2},& if \
$|\lambda | < 2 \sigma$;\cr 0,&otherwise.} 
\label{eq_semicircle}
\eea

This theorem is also known as 
{\it Wigner's law} \cite{Wigner},
and its extensions to further matrix ensembles 
have long been used for the stochastic treatment of 
complex quantum mechanical systems lying far beyond the
reach of exact methods \cite{Dyson,Wigner2}. Later,
the semi-circle law was found to have 
many applications in statistical physics and solid-state
physics as well \cite{Mehta_book,Vulpiani,Guhr}.

\par
\begin{figure}
\begin{center}
\begin{picture}(9,6.2)
\put(0.2,6.2){\psfig{width=6\unitlength, 
	angle=-90, file=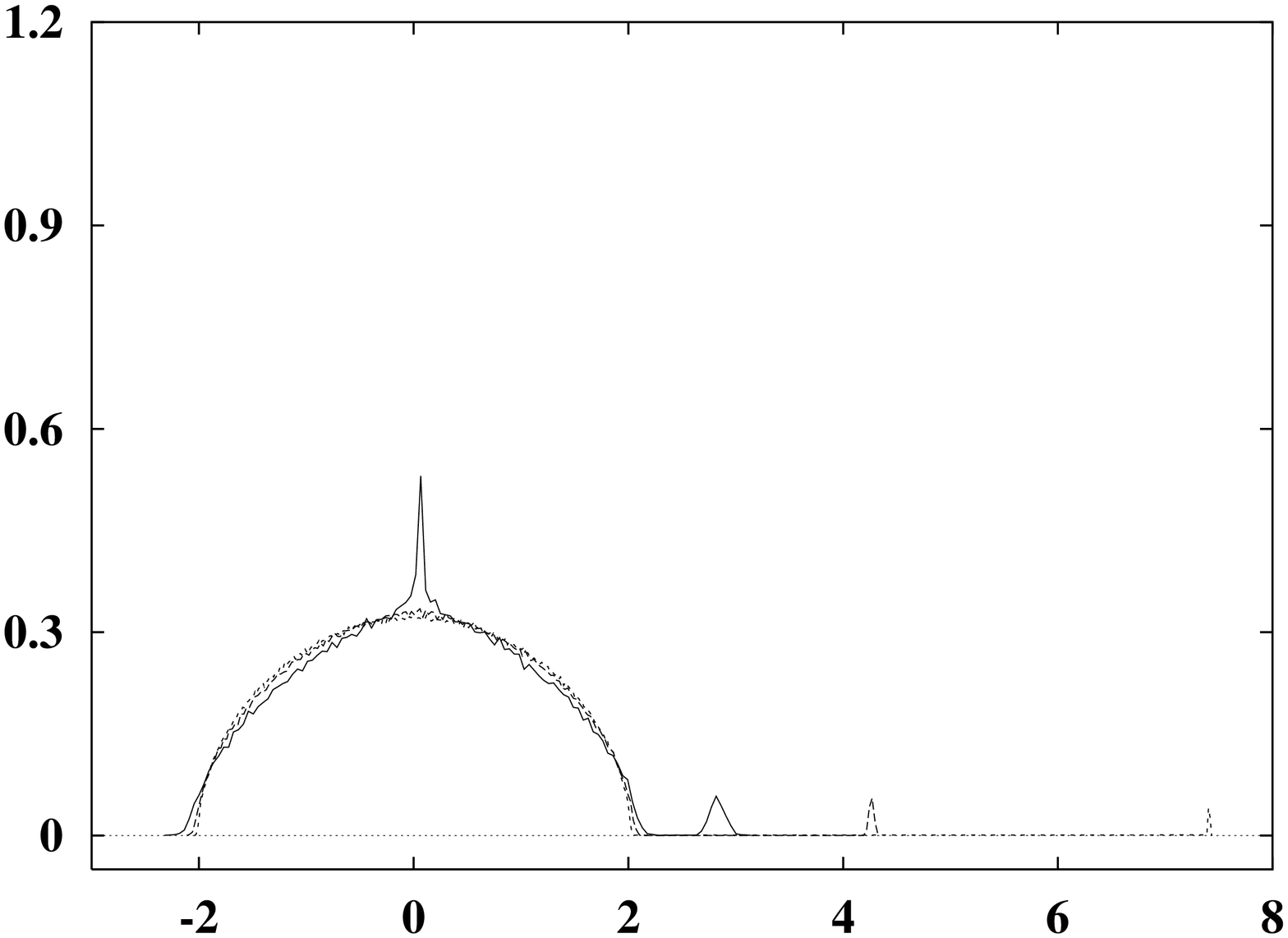}}
\put(-0.2,2.7){\psfig{width=2\unitlength,angle=90,file=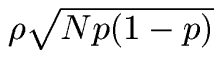}}
\put(3.5,0){$\lambda / \sqrt{Np(1-p)}$}
\put(4.8,5.8){\psfig{width=2.5\unitlength, angle=-90, 
file=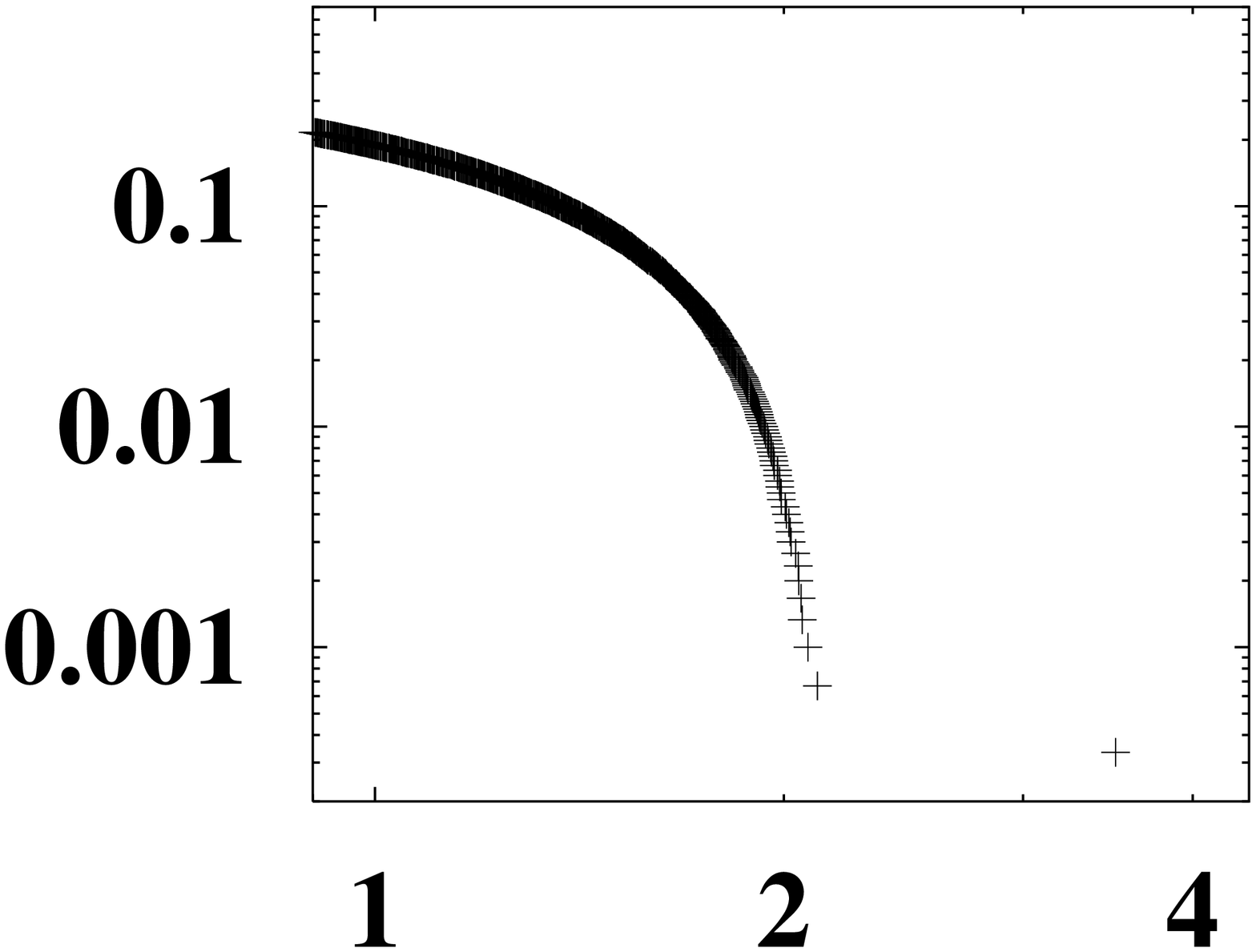}}
\put(6.4,3.1){\tiny ${\mathbf{\lambda\over\sqrt{Np(1-p)}}}$}
\put(4.9,5.4){\tiny ${\mathbf{1-F}}$}
\end{picture}
\end{center}
\caption[]{
If $N\to\infty$ and $p={\rm const}$, 
the average spectral density of an uncorrelated random graph
converges to a semi-circle, the first eigenvalue grows as $N$, and the
second is proportional to $\sqrt{N}$
(see Section\,\ref{ss_uncorr}). 
{\bf Main panel:}
The spectral density is shown for 
$p=0.05$ and three different system sizes: 
$N=100$ (---), $N=300$ \hbox{(-- --)} 
and $N=1000$ \hbox{(- - -)}.
In all three cases, the complete spectrum of $1000$ graphs 
was computed and averaged.
{\bf Inset:}
At the edge of the semi-circle 
-- i.e., in the 
$\lambda\approx\pm 2\sqrt{Np(1-p)}$ regions, --
the spectral density decays exponentially,
and with $N\to\infty$, the decay rate diverges\cite{Mehta_book,Bronx}.
Here, $F(\lambda)= N^{-1} \sum_{\lambda_i<\lambda} 1$ 
is the cumulative spectral distribution function,
and $1-F$ is shown 
for a graph with $N=3,000$ vertices and $15,000$ edges.
\label{fig_ER}
}
\end{figure}

Note, that for the adjacency matrix of the
uncorrelated random graph
many of the semi-circle law's conditions do not hold,
e.g., the expectation value of
the entries is a non-zero constant: $p\not= 0$.
Nevertheless, in the $N\to\infty$ limit,
{\it the rescaled spectral density of
the uncorrelated random graph
converges to the semi-circle law} 
of Eq.\,(\ref{eq_semicircle})\cite{Juhasz}.
An illustration of the convergence 
of the average spectral density to the semi-circular
distribution can be seen on Fig.\,\ref{fig_ER}. 
It is necessary to make a comment concerning figures here.
In order to keep figures simple, for the spectral density plots
we have chosen to show the spectral density of the original matrix,
$A$, and to rescale the horizontal ($\lambda$) 
and vertical ($\rho$) axes by 
$\sigma^{-1} N^{-1/2}=[Np(1-p)]^{-1/2}$ 
and $\sigma N^{1/2}=[Np(1-p)]^{1/2}$.

\unitlength10mm 

Some further results on the behavior of the uncorrelated
random graph's eigenvalues, 
relevant for the analysis of real-world graphs as well, 
include the following:
{\it The principal eigenvalue} (the largest eigenvalue,
$\lambda_1$) {\it grows much faster than the second eigenvalue}:
$\lim_{N\to\infty}(\lambda_1 / N) = p$
with probability 1, 
whereas for every $\epsilon > 1/2$, 
$\lim_{N\to\infty}(\lambda_2 / N^{\epsilon} ) = 0$.
(see Refs.\,\cite{Juhasz,CvetRowl} and Fig.\,\ref{fig_ER}.).
A similar relation holds for the smallest eigenvalue, 
$\lambda_N$: 
for every $\epsilon > 1/2$, 
$\lim_{N\to\infty}(\lambda_N / N^{\epsilon} ) = 0$. 
In other words: if $\langle k_i\rangle$ denotes the average number
of connections of a vertex in the graph, then $\lambda_1$ scales as 
$pN\approx\langle k_i\rangle$, 
and the width of the ``bulk'' part of the spectrum 
-- the set of the eigenvalues $\{\lambda_2, ...,\lambda_N\}$ --
scales as $\sigma\sqrt{N}$. 
Lastly, the semi-circular distribution's
edges are known to decay exponentially,
and the number of eigenvalues in the $\lambda >{\mathcal O} (\sqrt{N})$
tail has been shown to be of 
the order of $1$\cite{Mehta_book,Bronx}.

\subsection{Real-world graphs}
\label{ss_real}

The two main models proposed to describe real-world
graphs are the 
{\it small-world model} and the {\it scale-free model}.

\subsubsection{Small-world graphs}
\label{sss_sw}

The {\it small-world graph}\cite{WattsStrogatz,Watts_book,Newman_meanfield} 
is created 
by randomly rewiring some of
the edges of a regular\cite{def_regular} ring graph.
The {\it regular ring graph} is created as follows.
First draw the vertices $1$, $2$, ..., $N$
on a circle in ascending order.
Then, for every 
$i$, connect vertex $i$ to the vertices
lying closest to it on the circle:
vertices $i-k/2$, ..., $i-1$, $i+1$, ..., $i+k/2$, 
where every number should be understood modulo $N$
($k$ is an even number).
Fig.\,\ref{fig_SW.symmetries} shows, that this algorithm 
creates a regular graph indeed, 
because the degree\cite{def_regular}
of any vertex is the same number, $k$.
Next, starting from vertex $1$ and
proceeding towards $N$, perform the
{\it rewiring step}. For vertex $1$, 
consider the first ``forward connection'', i.e., the
connection to vertex $2$. 
With probability $p_{\rm r}$, reconnect vertex $1$ to another vertex
chosen uniformly at random and without allowing multiple edges. 
Proceed towards the remaining forward connections of vertex $1$,
and then, perform this step for the remaining $N-1$ vertices also.
For the rewiring, use equal and independent probabilities.
Note, that in the small-world model the density of edges is
$p=\langle k_i\rangle /(N-1)\approx k/N$.
Throughout this paper, we will use only $k>2$.

\par
If we use $p_{\rm r}=0$ in the small-world model,
the original regular graph is preserved, and for $p_{\rm r}=1$, 
one obtains a random graph
which differs from the uncorrelated random graph only slightly:
every vertex has a minimum degree of $k/2$.
Next, we will need two definitions. 
The {\it separation} between 
vertices $i$ and $j$ -- denoted by $L_{ij}$ --
is the number of edges in the shortest path 
connecting them.
The {\it clustering coefficient} 
at vertex $i$ -- denoted by $C_i$ --
is the number of existing
edges among the neighbors of vertex $i$ 
divided by the number of all possible
connections between them. 
In the small-world model, both $L_{ij}$ and $C_i$ are 
functions of the rewiring probability, $p_{\rm r}$.
Based on the above definitions of $L_{ij}(p_{\rm r})$
and $C_i(p_{\rm r})$,
the characteristics of the 
{\it small-world phenomenon} 
-- which occurs for intermediate values of $p_{\rm r}$ --
can be given as follows\cite{WattsStrogatz,Watts_book}:
(i) the average separation between two vertices, $L(p_{\rm r})$, 
drops dramatically below $L(p_{\rm r}=0)$, whereas
(ii) the average clustering coefficient, 
$C(p_{\rm r})$, remains high, close to $C(p_{\rm r}=0)$.
Note, that the 
rewiring procedure is carried out independently for every edge,
therefore the degree sequence and also other distributions
in the system -- e.g., path length and loop size -- 
{\it decay exponentially}.

\subsubsection{The scale-free model}
\label{sss_sf}

The {\it scale-free model}
assumes a random graph to be a {\it growing} set of
vertices and edges, where the location of new edges is determined by  
a {\it preferential attachment rule}\cite{EmergenceOfScaling,ScaleFree_PA}.
Starting from an initial
set of $m_0$ isolated vertices, one adds $1$ new vertex and $m$ new
edges at every time step $t$. 
(Throughout this paper, we will use $m=m_0$.)
The $m$ new edges connect the new
vertex and $m$ different vertices chosen from the $N$ old vertices.
The $i$th old vertex is
chosen with probability $k_i / \sum_{j=1,N} k_j$, where $k_i$ is
the degree of vertex $i$. (The density of edges in a scale-free graph is 
$p=\langle k_i\rangle / (N-1)\approx 2m/N$.)
In contrast to the small-world model, 
the distribution of degrees in a scale-free graph
converges to a power-law when $N\to\infty$,
which has been shown to be 
a combined effect of growth and
the preferential attachment\cite{ScaleFree_PA}. 
Thus, in the infinite time or size limit, 
{\it the scale-free model has no characteristic scale in the degree size}
\cite{Newman_collab,Redner2,Redner3,Dorog_aging,Dorog_herit,Dorog_cont,Reka_thesis}.

\subsubsection{Related models}
\label{sss_related}

Lately,
numerous other models have been suggested for a 
{\it unified description} of real-world graphs
\cite{Newman_collab,Redner2,Redner3,Dorog_aging,Dorog_herit,Reka_thesis,BoseEinstein,Ginestra_Multiscale,Vaz}.
Models of growing networks with aging vertices 
were found to display both heavy tailed and
exponentially decaying degree sequences
\cite{Dorog_aging,Dorog_herit,Dorog_cont} 
as a function of the speed of aging.
Generalized preferential attachment rules have helped us better
understand the origin of the
exponents and correlations emerging in these 
systems\cite{Redner2,Redner3}.
Also, investigations of more complex network models
-- using aging or an additional fixed cost of edges 
\cite{Amaral_ClassesOfSW} 
or preferential growth and random rewiring
\cite{Reka_thesis} --
have shown, 
that in the ``frequent rewiring, fast aging, high cost'' 
limiting case, one obtains a graph with an
exponentially decaying degree sequence, whereas in the 
``no rewiring, no aging, zero cost'' limiting case 
the degree sequence will decay as a power-law.
According to studies of scientific collaboration networks 
\cite{Barab_collab,Newman_collab}
and further social and biological structures
\cite{Amaral_ClassesOfSW,WagnerFell,SoleMontoya},
a significant proportion of large networks 
lies between the two extremes. In such cases, the
{\it characterization} of the system using a small number of
{\it algebraic constants} could facilitate 
the {\it classification of real-world networks}.

\section{Tools}
\label{s_tools}

\subsection{Analytical}
\label{ss_anal}

\subsubsection{The spectrum of the graph}
\label{sss_spectrum}

{\it The spectrum of a graph is the set of eigenvalues of the graph's
adjacency matrix}. The physical meaning of a graph's eigenpair
(an eigenvector and its eigenvalue) can be illustrated by the
following example. Write each component of a vector $\vvec$ on
the corresponding vertex of the graph: $v_i$ on vertex $i$. 
Next, on every vertex write the sum of the numbers found on
the neighbors of vertex $i$. 
If the resulting vector is a multiple of $\vvec$, then $\vvec$ is an
eigenvector, and the multiplier is the corresponding eigenvalue of the
graph.

\par
{\it The spectral density of a graph is the 
density of the eigenvalues of its adjacency matrix}.
For a finite system, this can be written as a sum of delta functions

\bea
\rho(\lambda) := {1\over N} \sum_{j=1}^N 
\delta(\lambda -\lambda_j) \, ,
\label{eq_specdens}
\eea

\noindent 
which converges to a continuous function with $N\to\infty$
($\lambda_j$ is the $j$th largest eigenvalue of the graph's
adjacency matrix).

\par 
The spectral density of a graph can be
{\it directly related to the graph's topological features:}
the $k$th moment, $M_k$, of $\rho(\lambda)$ can be written as

\bea
M_k =
{1\over N} \sum_{j=1}^N (\lambda_j)^k =
{1\over N} {\rm Tr} \big( A^k\big) =\nonumber\\
={1\over N}
\sum_{i_1,i_2,\cdots ,i_k}
A_{i_1,i_2} A_{i_2,i_3} \cdots A_{i_k,i_1} \, .
\label{eq_mk}
\eea

From the topological point of view,
$D_k=NM_k$ is the {\it number of directed paths} (loops) 
of the underlying -- undirected -- graph, that return
to their starting vertex after $k$ steps.
On a tree, the length of any such 
path can be an even number only, because these paths 
contain any edge an even number of times: once such a path
has left its starting point by chosing a starting edge,
no alternative route for returning to the starting point
is available. However, if the graph contains loops of odd length, 
the path length can be an odd number, as well.


\subsubsection{Extremal eigenvalues}
\label{sss_ee}

In an uncorrelated random graph
the {\it principal eigenvalue}, $\lambda_1$, shows the density of edges
and $\lambda_2$ can be related to the
conductance of the graph as a network of resistances\cite{CombHandbook}. 
An important property of all graphs is the following: 
the principal eigenvector, $\evec_1$, of the adjacency
matrix is a non-negative vector (all components are non-negative), 
and if the graph has no isolated vertices, $\evec_1$ is a 
positive vector\cite{CvetDooSa}. All other eigenvectors are
orthogonal to $\evec_1$, therefore they all have entries with mixed signs.

%
%




\subsubsection{The inverse participation ratios of eigenvectors}
\label{sss_ipr}

The {\it inverse participation ratio} 
of the normalized $j$th eigenvector,
$\evec_j$, is defined as\cite{Guhr} 

\bea
I_j = \sum_{k=1}^N \big[(e_j)_k\big]^4 \, .
\eea

If the components of an eigenvector are identical 
-- $(e_j)_i = 1/\sqrt{N}$ for every $i$, -- 
then $I_j=1/N$. 
For an eigenvector with one single non-zero component
-- $(e_j)_i = \delta_{i,i'}$ -- 
the inverse participation ratio is $1$.
The comparison of these two extremal cases illustrates that 
with the help of
the inverse participation ratio, one can tell whether 
only ${\mathcal O}(1)$ or as many as ${\mathcal O}(N)$  
components of an eigenvector differ
significantly from $0$, i.e., whether an
{\it eigenvector is localized or non-localized}.

\subsection{Numerical}
\label{ss_num}

\subsubsection{General real symmetric eigenvalue solver}
\label{sss_general}

To compute the eigenpairs of graphs {\it below the size $N=5,000$}, 
we used the general real symmetric eigenvalue solver of
Ref. \cite{numrec_tred.tqli}. This algorithm requires the 
allocation of memory space to all entries of the matrix, thus
to compute the spectrum of a graph of size $N=20,000$ 
($N=1,000,000$) using this general method with
double precision floating point arithmetic,
one would need $3.2$~GB ($8$~TB) memory space and 
the execution of approximately
$30 ~N^2 = 1.2\times 10^{10}$
($3\times 10^{13}$) floating point operations
\cite{numrec_tred.tqli}.
Consequently, we need to develop more
efficient algorithms to investigate the properties of graphs 
with sizes comparable to real-world networks.

\subsubsection{Iterative eigenvalue solver based on the thick-restart Lanczos algorithm}
\label{sss_iter_eigsolv}

The spectrum of a real-world 
graph is the spectrum of a sparse real symmetric
matrix, therefore the most efficient algorithms which can give 
a handful of the top $n_{\rm d}$ eigenvalues 
-- and the corresponding eigenvectors --
of a large graph are iterative methods\cite{Parlett}. 
These methods allow the matrix to be stored in any compact format, as
long as matrix-vector multiplication can be carried out at a high speed. 
{\it Iterative methods use little memory}: 
only the non-zero entries of the 
matrix and a few vectors of size $N$ need to be stored. 
The price for computational speed lies in the number of
the obtained eigenvalues: iterative methods {\it compute only a
handful of the largest} (or smallest) {\it eigenvalues} of a matrix.
To compute the eigenvalues of graphs above the size $N=5,000$, 
we have developed algorithms using a specially modified
version of
the thick-restart Lanczos algorithm\cite{Wu_JCP,Wu_LBNL41412}.
The modifications and some of the main technical parameters of
our software are explained in the following paragraphs.

\par
Even though
iterative eigenvalue methods 
are mostly used to obtain the top eigenvalues of a
matrix, after minor modifications the 
{\it internal eigenvalues in the vicinity of a fixed 
$\lambda=\lambda_0$ point} can be computed as well. For this, 
extremely sparse matrices are usually
``shift-inverted'', i.e., to find those eigenvalues of 
$A$ that are closest to $\lambda_0$, the highest and lowest
eigenvalues of $(A-\lambda_0 I)^{-1}$ are searched for. However,
because of the extremely high cost of matrix inversion 
in our case, for the computation of internal eigenvalues
we suggest to use the {\it ``shift-square'' method} with the matrix

\bea
B = \big[\lambda^{*} / 2 - (A-\lambda_0 I)^2\big]^{2n+1} \, .
\label{eq_Lanz}
\eea

\noindent
Here $\lambda^{*}$ is the largest
eigenvalue of $(A-\lambda_0 I)^2$,
$I$ is the identity matrix and $n$ is a positive
integer. Transforming the matrix $A$ into $B$ transforms the spectrum
of $A$ in the following manner. 
First, the spectrum is shifted to the left by $\lambda_0$.
Then, the spectrum is ``folded'' (and squared) at the origin
such that all eigenvalues will be negative. Next, the spectrum is
linearly rescaled and shifted to the right, with the following effect:
(i) the whole spectrum will lie in the symmetric interval 
$[-\lambda^{*}/2,\lambda^{*} / 2]$ and
(ii) those eigenvalues that were closest to $\lambda_0$ in the
spectrum of $A$, will be the largest now, i.e.,
they will be the eigenvalues
closest to $\lambda^{*} / 2$.
Now, raising all eigenvalues to the $(2n+1)$st power increases the
relative difference, $1-\lambda_i/\lambda_j$, 
between the top eigenvalues $\lambda_i$ and $\lambda_j$ 
by a factor of $2n+1$. 
This allows the iterative method find the top eigenvalues of $B$ more quickly.
One can compute the {\it corresponding
eigenvalues} (those being closest to $\lambda_0$) 
{\it of the original matrix}, $A$:
if $\vec{b}_1$, $\vec{b}_2$, ..., $\vec{b}_{n_{\rm d}}$ 
are the normalized eigenvectors
of the $n_{\rm d}$ largest eigenvalues of $B$, 
then for $A$ the $n_{\rm d}$
eigenvalues closest to $\lambda_0$ will be -- not necessarily in
ascending order --
$\vec{b}_1\, A\,\vec{b}_1$, $\vec{b}_2\, A\,\vec{b}_2$, ..., 
$\vec{b}_{n_{\rm d}}\, A\,\vec{b}_{n_{\rm d}}$.

\par
The thick-restart Lanczos method
uses memory space for the non-zero entries of the
$N\times N$ large adjacency matrix,
and $n_{\rm g}+1$ vectors of length $N$, 
where $n_{\rm g}$ ($n_{\rm g} > n_{\rm d}$) 
is usually between $10$ and $100$.
Besides the relatively small size of required memory,
we could also exploit the fact that the 
{\it non-zero entries of a graph's adjacency matrix are all 1's}:
during matrix-vector multiplication 
-- which is usually the most time-consuming
step of an iterative method --
only additions had to be carried out instead of multiplications.

\par
The {\it numerical spectral density functions} 
of large graphs 
($N\ge 5,000$) of this paper were
obtained using the following steps.
To compute the spectral density of the adjacency matrix, $A$,
at an internal $\lambda=\lambda_0$ location,
first, the $n_{\rm d}$ eigenvalues closest to $\lambda_0$ were searched for.
Next, the distance between the smallest and the largest of the
obtained eigenvalues was computed.  
Finally, $\rho(\lambda_0)$ this distance was multiplied by $N/(n_{\rm d}-1)$,
and was averaged using $n_{\rm av}$ different graphs.
We used double precision floating point
arithmetic, and the iterations were stopped, if (i) at least 
$n_{\rm it}$ iterations had been carried out 
and (ii) the lengths of the residual vectors
belonging to the $n_{\rm d}$ selected eigenpairs 
were all below $\varepsilon = 10^{-12}$\cite{Wu_JCP}.

\section{Results}
\label{s_res}

\subsection{Sparse uncorrelated random graphs: the semi-circle law is
not universal}
\label{ss_ER.sparse}

In the uncorrelated random graph model of 
Erd\H os and R\'enyi, the total number of edges
grows quadratically with the number of vertices:
$N_{\rm edge}=N \langle k_i\rangle = N p(N-1)\approx pN^2$. 
However, {\it in many real-world graphs edges are ``expensive''},
and the growth rate of the number of 
connections remains well below this rate.
For this reason, we also investigated the spectra of such uncorrelated
networks, for which the probability of any two vertices being connected
changes with the size of the system using $pN^{\alpha}=c={\rm const}$. 
Two special cases are $\alpha=0$ (the Erd\H os-R\'enyi model) and $\alpha=1$.
In the second case, $pN\to {\rm const}$ as $N\to\infty$, i.e.,
the average degree remains constant.

\par
For {\it $\alpha<1$ and $N\to\infty$},
there exists an infinite cluster of connected vertices
(in fact, it exists for every $\alpha\le 1$ \cite{Bollobas}).
Moreover, the expectation value of any
$k_i$ converges to infinity, thus any vertex is
almost surely connected to the infinite cluster.
The {\it spectral density
function converges to the semicircular distribution} of 
Eq.\,(\ref{eq_semicircle}), because the total weight of
isolated subgraphs decreases exponentially 
with growing system size.
(A detailed analysis of this issue is available in 
Ref.\,\cite{BollobasThomason85}.)

\par
\begin{figure}
\begin{center}
\begin{picture}(9,6.2)
\put(0,6.2){\psfig{width=6\unitlength, 
	angle=-90, file=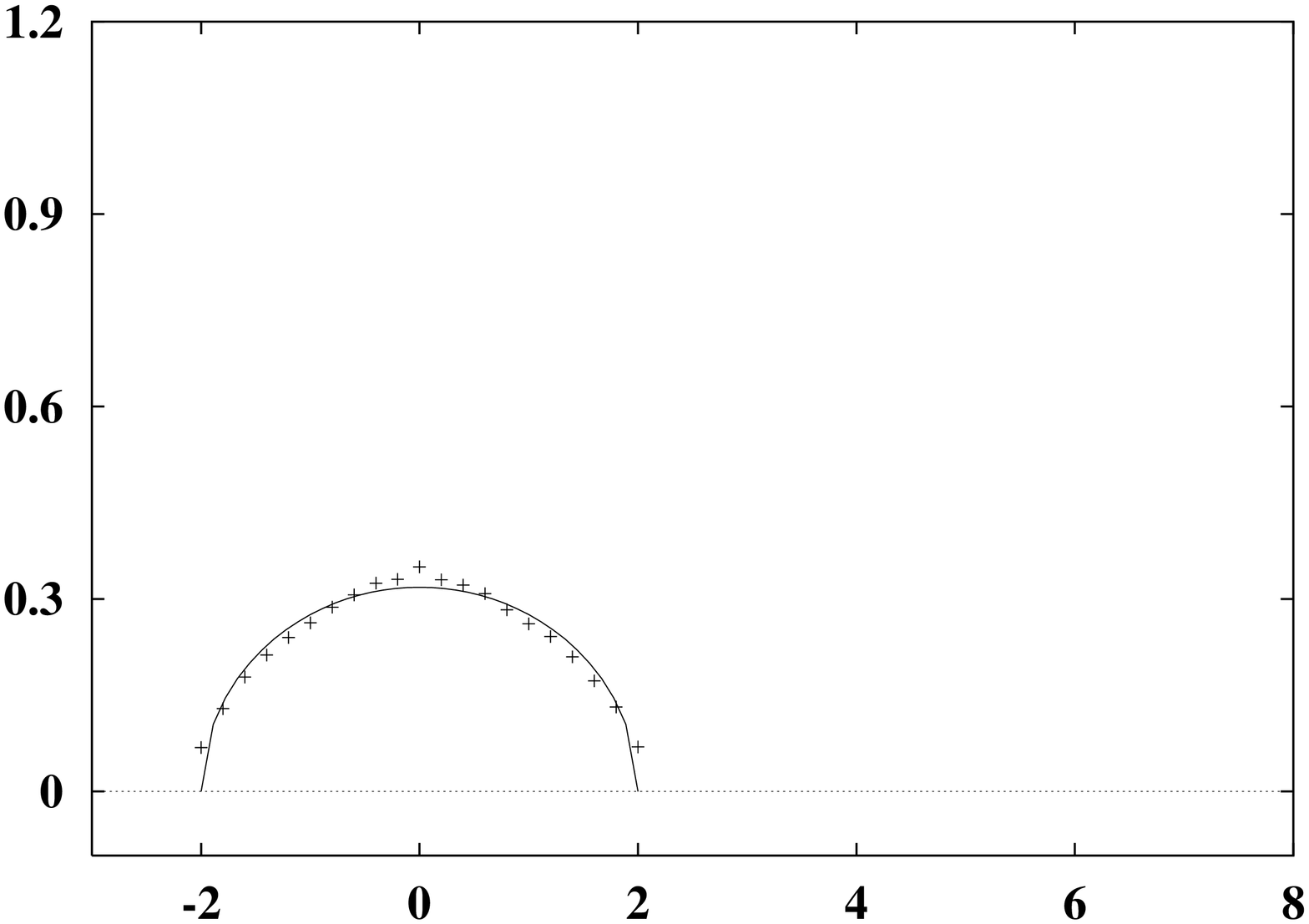}}
\put(-0.2,2.7){\psfig{width=2\unitlength,angle=90,file=rhoNp1p.ps}}
\put(4,0){$\lambda / \sqrt{Np(1-p)}$}
\put(4.3,5.9){\psfig{width=2.8\unitlength, angle=-90, 
	file=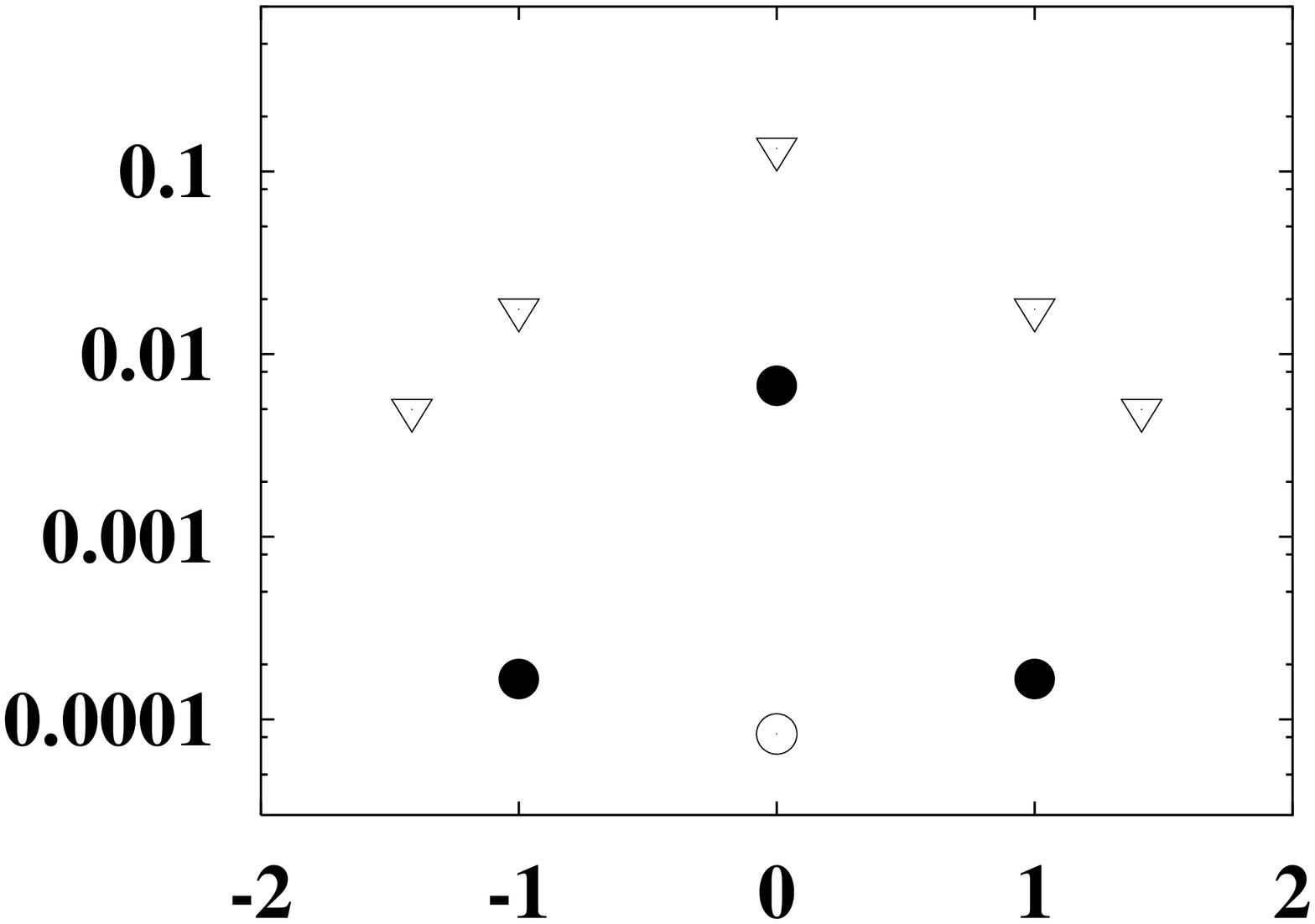}}
\put(4,5.6){\tiny ${\mathbf{{\rm strength\,of\,}\delta}}$}
\put(7,3.1){\tiny ${\mathbf{\lambda}}$}
\end{picture}
\end{center}
\caption[]{
If $N\to\infty$ and $pN={\rm const}$, the spectral density 
of the uncorrelated random graph does not 
converge to a semi-circle. 
{\bf Main panel:} 
Symbols show the spectrum of an uncorrelated random graph 
($20,000$ vertices and $100,000$ edges)
measured with the iterative method 
using $n_{\rm av}=1$, $n_{\rm d}=101$ and $n_{\rm g}=250$.
A solid line shows the semi-circular distribution for comparison.
(Note, that the principal eigenvalue, $\lambda_1$, is not shown here,
because here at any $\lambda_0$ point
the average first neighbor distance among $n_{\rm d}=101$ 
eigenvalues was used to measure the spectral density.)
{\bf Inset:} 
Strength of delta functions in $\rho(\lambda)$ 
``caused'' by isolated clusters of sizes $1$, $2$ and $3$
in uncorrelated random graphs 
(see Ref. \cite{delta} for a detailed explanation).
Symbols are for graphs with $20,000$ vertices and
$20,000$ edges ($\nabla$), 
$50,000$ edges ($\bullet$), 
and $100,000$ edges ($\circ$).
Results were averaged for $3$ different graphs everywhere.
\label{fig_ERsparse}
}
\end{figure}

For {\it $\alpha=1$ and $N\to\infty$}
(see Fig.\,\ref{fig_ERsparse}),
the probability
for a vertex to belong to a cluster of any finite size remains
also finite \cite{alpha=1}. 
Therefore, the limiting spectral density 
contains the weighted sum of the spectral densities of 
all finite graphs\cite{delta}. 
The most striking deviation from the semi-circle law in this case is
the {\it elevated central part} of the spectral density.
The probability for a vertex to
belong to an isolated cluster
of size $s$ decreases exponentially with $s$\cite{alpha=1},
therefore the number of large isolated clusters is low
and the eigenvalues of a graph with $s$ vertices are bounded by
$-\sqrt{s-1}$ and $\sqrt{s-1}$. For these two reasons, 
the amplitudes of delta functions 
decay exponentially, as the absolute value of their locations,
$|\lambda|$, increases. 

\par
The principal eigenvalue of this graph
converges to a constant: $\lim_{N\to\infty}(\lambda_1)=pN=c$, 
and $\rho(\lambda)$ will be symmetric in the $N\to\infty$
limit.
Therefore, in the limit, all odd moments ($M_{2k+1}$), and thus, the
number of all loops with odd length ($D_{2k+1}$) disappear. This is a
salient feature of graphs with tree structure (because on a tree every
edge must be used an even number of times in order to return to the
initial vertex), indicating that the structure of a sparse uncorrelated
random graph becomes more and more tree-like. This can also be
understood by considering that the typical distance (length of the
shortest path) between two vertices on both a sparse uncorrelated
random graph and a regular tree with the same number of edges scales as
$\ln(N)$. 
So {\it except for a few short-cuts
a sparse uncorrelated random
graph looks like a tree}.


\subsection{The small-world graph}
\label{ss_SW}

\subsubsection{Triangles are abundant in the graph}
\label{sss_tri}

\begin{figure}
\begin{center}
\begin{picture}(9,6.2)
\put(-0.1,6.6){\psfig{width=3.2\unitlength, angle=-90, file=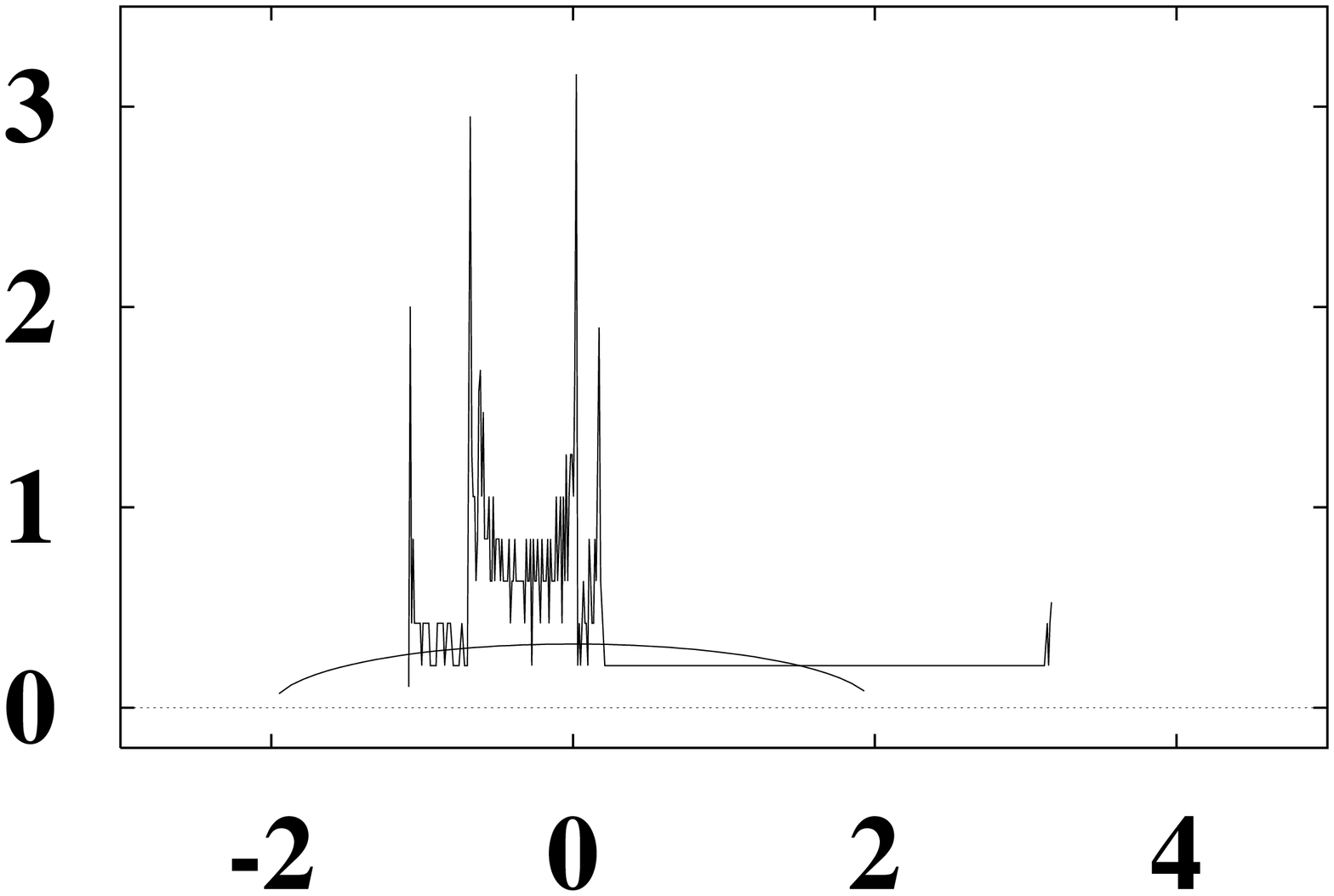}}
\put(3.7,5.9){\tt a}
\put(4,6.6){\psfig{width=3.2\unitlength, angle=-90, file=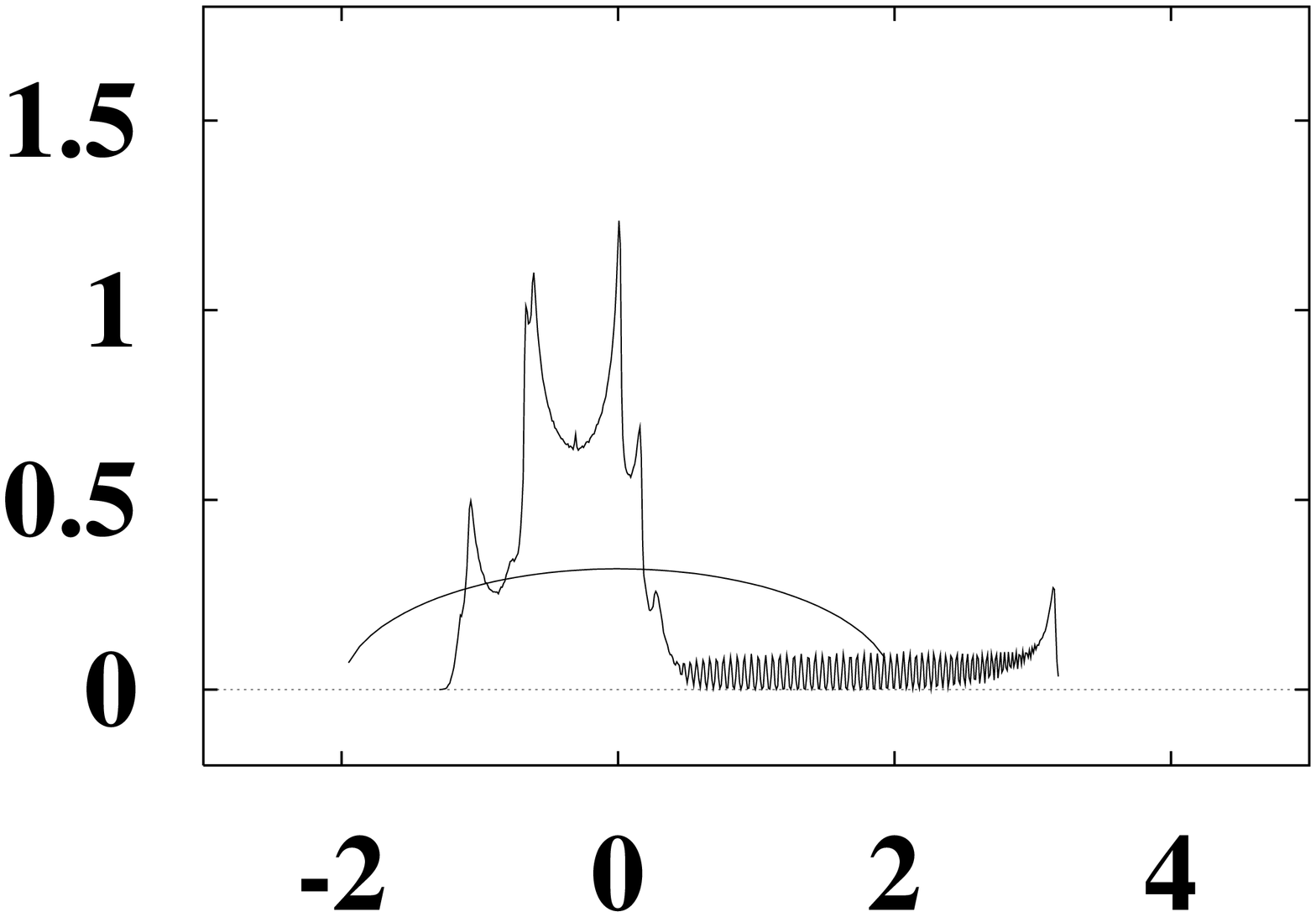}}
\put(7.9,5.9){\tt b}
\put(-0.1,3.6){\psfig{width=3.2\unitlength, angle=-90, file=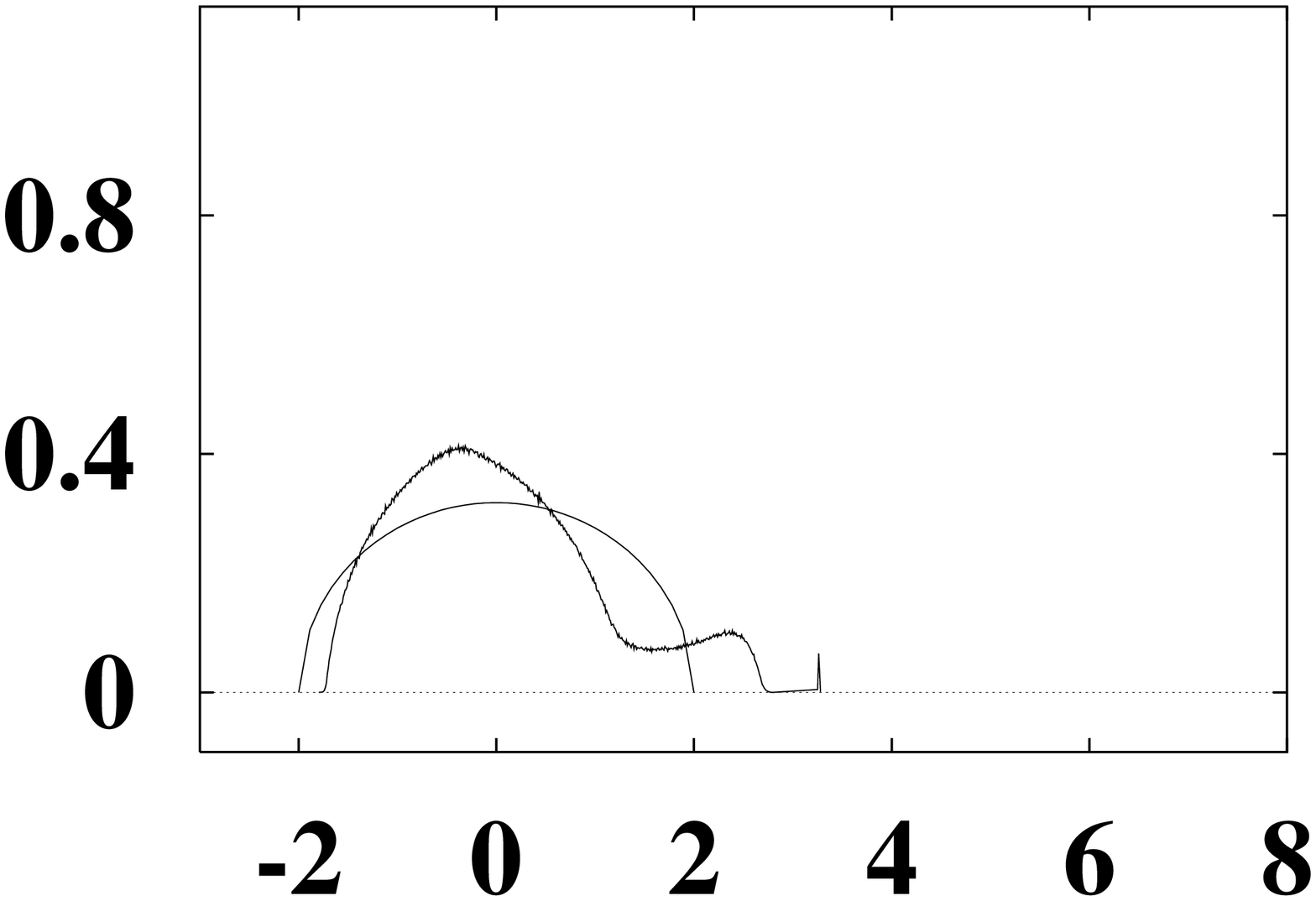}}
\put(1.5,0.2){\small $\lambda\, [Np(1-p)]^{-1/2}$}
\put(3.7,2.9){\tt c}
\put(4,3.6){\psfig{width=3.2\unitlength, angle=-90, file=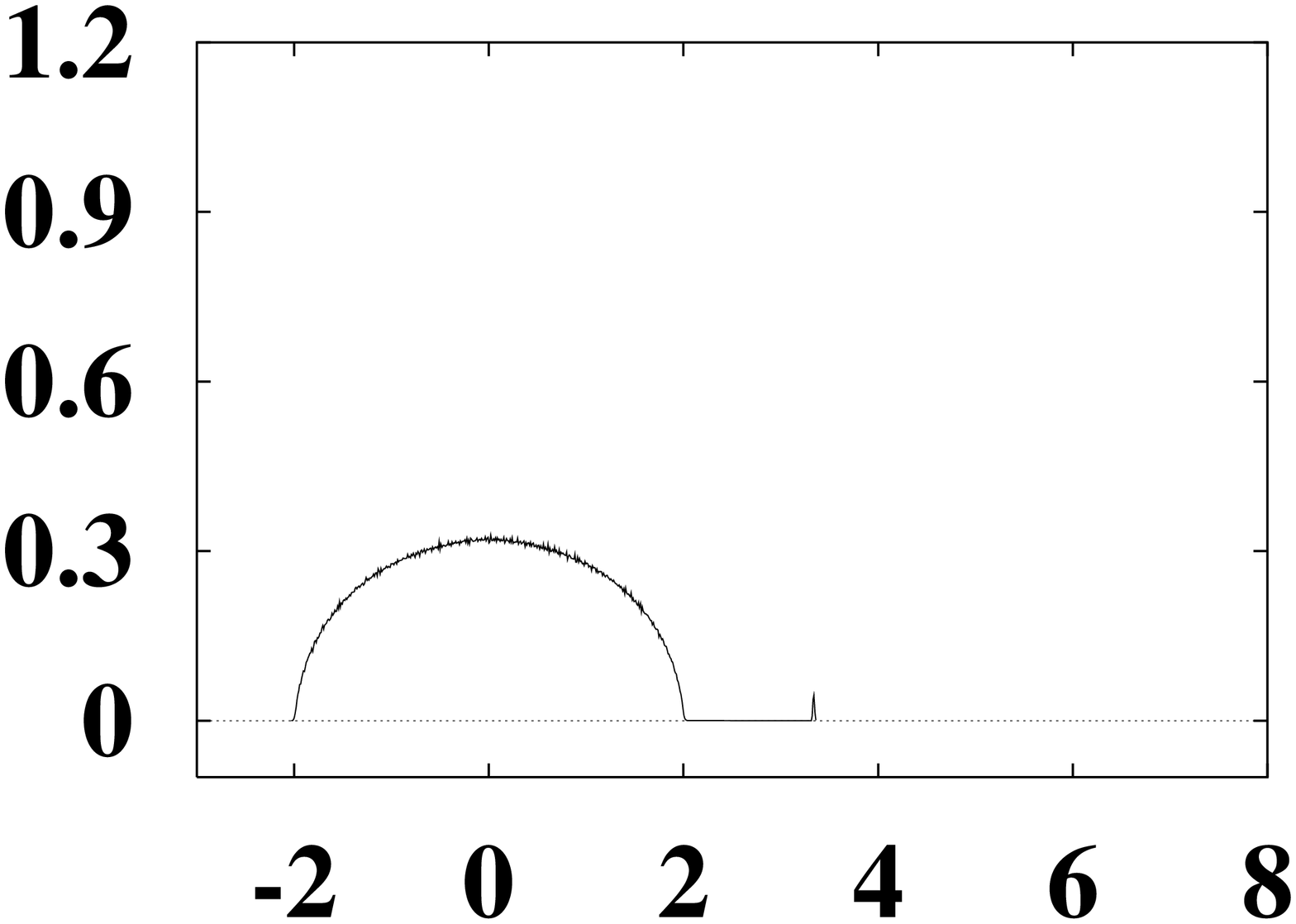}}
\put(5.7,0.2){\small $\lambda\, [Np(1-p)]^{-1/2}$}
\put(7.9,2.9){\tt d}
\put(-0.2,4.5){\psfig{width=1.5\unitlength,angle=90,file=rhoNp1p.ps}}
\put(-0.2,1.2){\psfig{width=1.5\unitlength,angle=90,file=rhoNp1p.ps}}
\end{picture}
\end{center}
\caption[]{
Spectral densities of small-world graphs using the complete spectra.
The solid line shows the semi-circular distribution for comparison.
{\bf (a)}
Spectral density of the
regular ring graph created from the small-world model
with $p_{\rm r}=0$, $k=10$ and $N=1000$. 
{\bf (b)} 
For $p_{\rm r}= 0.01$, the average spectral density of small-world graphs 
contains sharp maxima, which are the ``blurred''
remnants of the singularities of the $p_{\rm r}=0$ case.
Topologically, this means, that the graph is still almost regular, 
but it contains a small number of impurities. 
In other words, after a small perturbation, the system is no more degenerate.
{\bf (c)}
The average spectral density computed for the
$p_{\rm r} = 0.3$ case shows that
the third moment of $\rho(\lambda)$ is preserved even for very 
high values of $p_{\rm r}$,
where there is already no sign of any blurred 
singularity (i.e., regular structure). This means, that
even though all remaining regular islands 
have been destroyed already,
{\it triangles are still dominant}.
{\bf (d)}
If $p_{\rm r} = 1$, then
the spectral density of the small-world graph converges to a semi-circle.
In figures {\bf b}, {\bf c} and {\bf d},
$1000$ different graphs with $N=1000$ and $k=10$ were used for averaging.
\label{fig_SW.p}
}
\end{figure}

For $p_{\rm r}=0$, the small-world graph is 
{\it regular and also periodical}. 
Because of the highly ordered structure,
$\rho(\lambda)$ contains numerous singularities, which are listed
in Section\,\ref{app_SW} (see also Fig.\,\ref{fig_SW.p}). 
Note that $\rho(\lambda)$ has a high third moment.
(Remember, that we use only $k>2$.)

\par
If we increase $p_{\rm r}$ such that
the small-world region is reached, 
i.e., the periodical structure of the graph is perturbed,
then singularities become blurred and are transformed 
into high local maxima, 
but $\rho(\lambda)$ retains a strong skewness (see Fig.~\ref{fig_SW.p}). 
This is in good agreement with results of 
Refs.~\cite{Newman_meanfield,Amaral_crossover},
where it has been shown that the local structure of the 
small-world graph is ordered, however, 
already {\it a very small number of shortcuts can 
drastically change the graph's global structure}.

\unitlength10mm 

In the $p_{\rm r}=1$ case
the small-world model becomes very similar to the uncorrelated
random graph: the only difference is that here,
the minimum degree of any vertex is a positive constant, $k/2$,
whereas in an uncorrelated random graph the degree of a vertex can be
any non-negative number.
Accordingly, $\rho(\lambda)$ becomes a
semi-circle for $p_{\rm r}=1$ (Fig.~\ref{fig_SW.p}). 
Nevertheless, it should be noted, 
that as $p_{\rm r}$ converges to $1$, 
a high value of $M_3$ is preserved even for 
$p_{\rm r}$ close to $1$, 
where all local maxima have already vanished.
The third moment of $\rho(\lambda)$ gives the number of triangles
in the graph (see Section\,\ref{sss_spectrum});
the lack of high local maxima 
-- i.e., the remnants of singularities -- 
shows the absence of an ordered structure.

\par
From the above we conclude, that 
-- from the spectrum's point of view --
the {\it high number of triangles
is one of the most basic properties of the small-world model},
and it is preserved much longer, than regularity or periodicity,
if the level of randomness, $p_{\rm r}$, is increased.
This is in good agreement with the results of Ref.~\cite{WagnerFell},
where the high number of small cycles is found to be a fundamental
property of small-world networks. As an application, the high number of
small cycles results in special diffusion properties on small-world 
graphs~\cite{relax}.

\subsection{The scale-free graph}
\label{ss_SF}

\begin{figure}
\begin{center}
\begin{picture}(9,6.2)
\put(0.2,6.2){\psfig{width=6\unitlength, 
	angle=-90, file=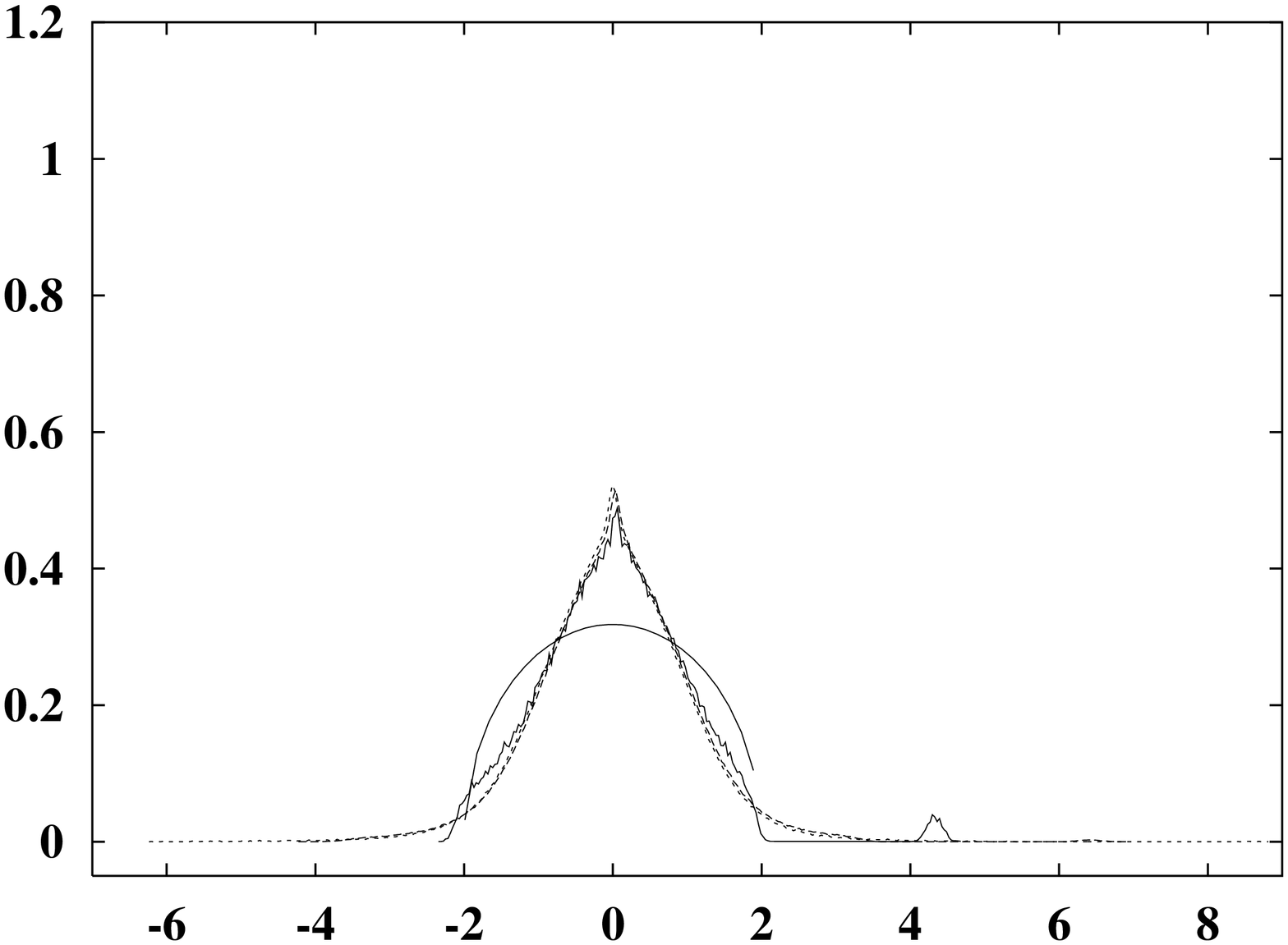}}
\put(-0.2,2.7){\psfig{width=2\unitlength,angle=90,file=rhoNp1p.ps}}
\put(5.5,0){$\lambda / \sqrt{Np(1-p)}$}
\put(4.1,6){\psfig{width=3.2\unitlength, angle=-90, 
	file=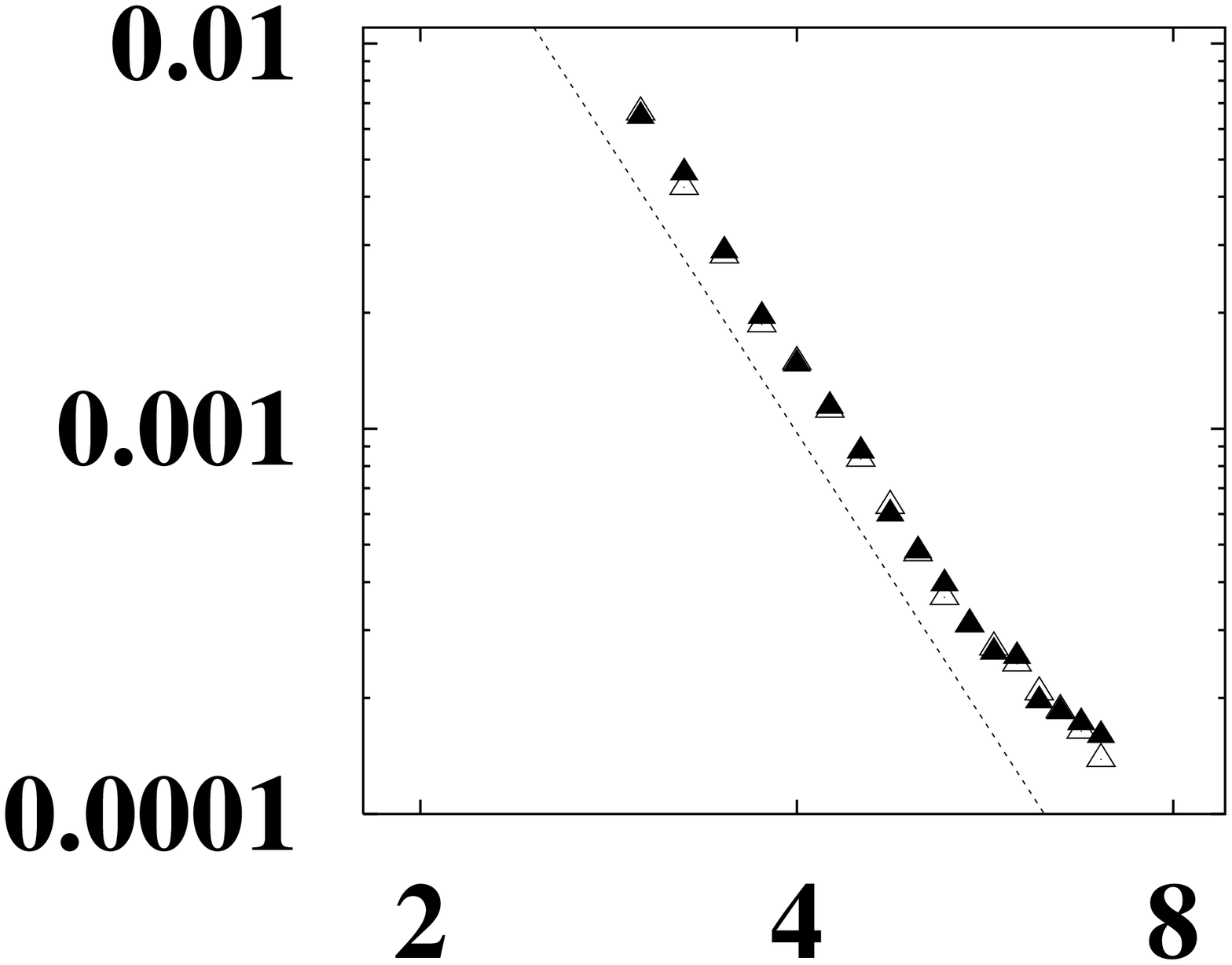}}
\put(4.2,3.7){\psfig{width=1.5\unitlength,angle=90,file=rhoNp1p.ps}}
\put(6.1,2.7){\tiny ${\mathbf{\lambda\, [Np(1-p)]^{-1/2}}}$}
\end{picture}
\end{center}
\caption[]{
{\bf Main panel:} The average spectral densities of 
scale-free graphs with $m=m_0=5$, and 
$N=100$ (---), $N=1000$ \hbox{(-- --)} 
and $N=7000$ \hbox{(- - -)} vertices.
(For $N=100$ and $N=1000$, the complete spectrum of $1000$ graphs,
and for $N=7000$, the complete spectrum of $25$ graphs was used.)
Another continuous line shows the semi-circular distribution 
for comparison. 
Observe that 
(i) the central part of the scale-free graph's 
spectral density is triangle-like, not semi-circular and 
(ii) the edges show a power-law decay, whereas the
the semi-circular distribution's edges decay exponentially,
i.e., it decays exponentially at the edges\cite{Mehta_book}.
{\bf Inset:} The upper and lower tails of $\rho(\lambda)$ 
(open and full triangles) 
for scale-free graphs with $N=40,000$ vertices 
the average degree of a
vertex being $\langle k_i\rangle = 2m = 10$, as before.
Note that both axes are logarithmic and 
$\rho(\lambda)$ has a power-law tail with the same decay rate at both
ends of the spectrum.
Here, we used the iterative eigenvalue solver of 
Section\,\ref{sss_iter_eigsolv} with 
$n_{\rm d}=21$, $n_{\rm av}=3$ and $n_{\rm g}=60$.
The line with the slope $-5$ in this figure is a guide to the eye.
\label{fig_BA.all}
}
\end{figure}

{\it For $m=m_0=1$, the scale-free graph is a tree} by definition
and its spectrum is symmetric\cite{CvetDooSa}.
In the $m>1$ case
$\rho(\lambda)$ consists of several
well distinguishable parts (see Fig.\,\ref{fig_BA.all}). 
The ``bulk'' part of the spectral density
-- the set of the eigenvalues \{$\lambda_2, ...,\lambda_N$\} --
converges to a symmetric continuous function which 
has a triangle-like shape for the normalized $\lambda$ 
values up to $1.5$ and has power-law tails.

\unitlength10mm 

The {\it central part} of the spectral density
{\it converges to a triangle-like shape}
with its top lying well above the semi-circle.
Since the scale-free graph is fully connected by definition,
the increased number of eigenvalues with small magnitudes
cannot be accounted to isolated clusters, as before. 
As an explanation, we suggest, that the {\it eigenvectors} 
of these eigenvalues are {\it localized on
a small subset of the graph's vertices}.
(This idea is supported by the high inverse participation ratios of 
these eigenvectors, see Fig.\,\ref{fig_IPR}).

\subsubsection{The spectral density of the scale-free graph decays as a
power-law}
\label{sss_powerlaw}

The inset of Fig.\,\ref{fig_BA.all} shows the
{\it tail of the bulk part} of the spectral density
for a graph with $N=20,000$ vertices and 
$100,000$ edges (i.e., $pN=10$). 
Comparing this to the inset of Fig.\,\ref{fig_ER}
-- where the number of vertices and edges 
is the same as here, -- 
one can observe 
the {\it power law decay} at the edge of the bulk
part of $\rho(\lambda)$.
As shown later, in Section\,\ref{ss_test}, 
the power-law decay in this region is
caused by eigenvectors localized on vertices with the highest
degrees. The power law decay of the degree sequence 
-- i.e., the existence of very high degrees --
is, in turn, due to the preferential 
attachment rule of the scale-free model.

\subsubsection{The growth rate of the principal eigenvalue shows
a crossover in the level of correlations}
\label{sss_SF.spec.eigval}

Since the adjacency matrix of a graph is a 
non-negative symmetric matrix,
the graph's largest eigenvalue, $\lambda_1$, is also the largest in 
magnitude (see, e.g., Theorem 0.2 of Ref.~\cite{CvetDooSa}). 
Considering the effect of the adjacency matrix on the base
vectors $(b_i)_j =\delta_{ij}$ ($i=1,2,...,N$), 
it can be shown that 
a lower bound for $\lambda_1$ is given 
by the length of the longest row vector of the adjacency matrix, 
which is the square root of the graph's largest degree, $k_1$.
Knowing that the largest degree of a 
scale-free graph grows as $\sqrt{N}$ \cite{ScaleFree_PA},
one expects $\lambda_1$ to grow as 
$N^{1/4}$ for large enough systems.

\par
Fig.\,\ref{fig_BA.lambda1} shows 
a rescaled plot of the scale-free graph's largest eigenvalue
for different values of $m$. 
In this figure, $\lambda_1$ is compared to 
the length of the longest row vector, $\sqrt{k_1}$
on the ``natural scale'' of these values,
which is $\sqrt{m}N^{1/4}$\cite{ScaleFree_PA}.
It is clear that if $m>1$ and the system is small,
then through several decades 
(a) $\lambda_1$ is larger than $\sqrt{k_1}$
and 
(b) the growth rate of $\lambda_1$ 
is well below the expected rate of $N^{1/4}$.
In the $m=1$ case, and for large systems,
(a) the difference between $\lambda_1$ and $\sqrt{k_1}$ vanishes 
and 
(b) the growth rate of the principal eigenvalue 
will be maximal, too.
This {\it crossover} in the 
behavior of the scale-free graph's principal eigenvalue 
is a specific property of 
{\it sparse growing correlated graphs},
and it
is a result of the changing level of correlations between the
longest row vectors (see Section\,\ref{app_crossover}).

\subsubsection{Comparing the role of the principal eigenvalue in the scale-free
graph and the $\alpha=1$ uncorrelated random graph: a comparison of structures}
\label{sss_rpr}

\begin{figure}
\begin{center}
\begin{picture}(9,6.2)
\put(4.5,0){$N$}
\put(0,6.2){\psfig{width=6\unitlength, angle=-90, 
file=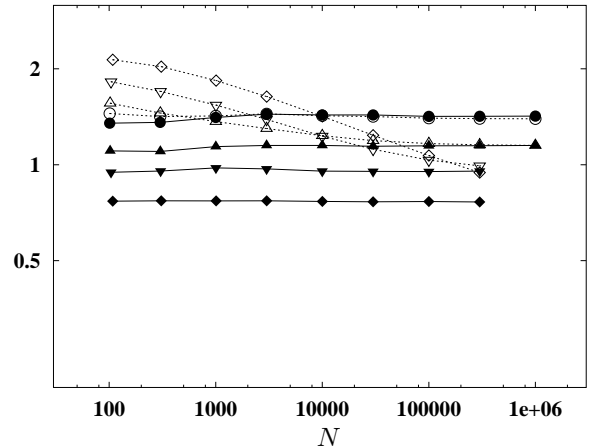}}
\end{picture}
\end{center}
\caption[]{
Comparison of the 
length of the longest row vector, $\sqrt{k_1}$, and 
the principal eigenvalue, $\lambda_1$, in scale-free graphs.
Open symbols show $\lambda_1\,/\,(\sqrt{m}N^{1/4})$,
closed symbols show $\sqrt{k_1}\,/\,(\sqrt{m}N^{1/4})$.
The parameter values are 
$m=1$ ($\circ$), 
$m=2$ ($\triangle$),
$m=4$ ($\nabla$) and
$m=8$ ($\diamond$).
Each data point is an average for $9$ graphs.
For the reader's convenience, data points are connected.
\\
If $m>1$ and the network is small,
the principal eigenvalue, $\lambda_1$, 
of a scale-free graph is determined by the 
largest row vectors jointly:
the largest eigenvalue is above $\sqrt{k_1}$ 
and the growth rate of $\lambda_1$
stays below the maximum possible growth rate, which is
$\lambda_1\propto N^{1/4}$.
If $m=1$, or the network is large, 
the effect of row vectors other than the longest on
$\lambda_1$ vanishes:
the principal eigenvalue 
converges to the length of the longest row vector,
and it grows as $\lambda_1\propto N^{1/4}$.
Our results show a {\it crossover in the growth rate
of the scale-free model's principal eigenvalue}.
\label{fig_BA.lambda1}
}
\end{figure}

Now we will compare the {\it role of the principal eigenvalue} 
in the $m>1$ scale-free graph and the
$\alpha=1$  uncorrelated random graph
through its effect on the moments of the spectral density.
On Figs.\,\ref{fig_BA.all} and \ref{fig_BA.lambda1}, one can observe that 
(i) the principal eigenvalue of the {\it scale-free graph} is detached from
the rest of the spectrum,
and (ii) as $N\to\infty$, it grows as $N^{1/4}$ 
(see also Sections \ref{sss_SF.spec.eigval} and \ref{app_crossover}).
The inset of Fig.\,\ref{fig_BA.all} indicates, that in the $N\to\infty$
limit, the bulk part will be symmetric, and its width will be constant
(Fig.\,\ref{fig_BA.all} rescales this constant width
merely by another constant, namely $Np[(1-p)]^{-1/2}$).
Because of the symmetry of the bulk part, in the $N\to\infty$ limit, 
the third moment of $\rho(\lambda)$
is determined exclusively by the contribution of the principal
eigenvalue, which is $N^{-1}(\lambda_1)^3\propto N^{-1/4}$. 
For each moment above the third (e.g., for the $l$th moment),
with growing $N$, 
the contribution of the bulk part to this moment will scale as 
${\mathcal O}(1)$, 
and the contribution of the principal eigenvalue will scale as 
$N^{-1+l/4}$. In summary, in the $N\to\infty$ limit,
the scale-free graph's first eigenvalue has a
significant contribution to the fourth moment;
the 5th and all higher moments are determined exclusively by
$\lambda_1$: the $l$th moment will scale as $N^{-1+l/4}$.

\par
In contrast to the above,
the principal eigenvalue of the $\alpha=1$ {\it uncorrelated random graph}
converges to the constant $pN=c$ in the $N\to\infty$ limit, and the
width of the bulk part also remains constant 
(see Fig.\,\ref{fig_ERsparse}). Given a fixed number, $l$, 
the contribution of the principal eigenvalue to the $l$th moment of
the spectral density will change as $N^{-1}c^l$ in the
$N\to\infty$ limit. The contribution of the bulk part will scale as
${\mathcal O}(1)$, 
therefore all even moments of the spectral density 
will scale as ${\mathcal O}(1)$ in
the $N\to\infty$ limit, and all odd moments will converge to $0$.

\par
The difference between
the growth rate of the moments of $\rho(\lambda)$ in the above two models 
(scale-free graph and $\alpha=1$ uncorrelated random graph model) 
can be interpreted as a sign of {\it different structure}
(see Section\,\ref{sss_spectrum}). 
In the $N\to\infty$ limit, the average degree of a vertex converges to a
constant in both models: 
$\lim_{N\to\infty} \langle k_i\rangle = pN = c = 2m$.
(Both graphs will have the same number of edges per vertex.)
On the other hand, in the limit,
all moments of the $\alpha=1$ uncorrelated random graph's spectral density
converge to a constant, whereas the moments $M_l (l=5,6,...)$ of the scale-free
graph's $\rho(\lambda)$ will diverge as $N{-1+l/4}$. 
In other words: the number of loops of length $l$ in the 
$\alpha=1$ uncorrelated random graph will grow as 
$D_l=NM_l={\mathcal O}(N)$, whereas for the scale-free graph for 
every $l\ge 3$, the number of these loops will grow as 
$D_l=NM_l={\mathcal O}(N^{l/4})$.
From this we conclude that in the
limit, the role of loops is negligible in the $\alpha=1$ uncorrelated
random graph, whereas it is large in the scale-free graph. In fact, 
the growth rate of the number of loops in the scale free graph exceeds
all polynomial growth rates:
the longer the loop size ($l$) investigated, the higher the growth
rate of the number of these loops ($N^{l/4}$) will be.
Note that the relative number of triangles (i.e., the third moment of
the spectral density, $M_l/N$) will disappear in the scale-free graph,
if $N\to\infty$.

\unitlength10mm 

In summary, the spectrum of the scale-free model converges to a
triangle-like shape in the center,
and the edges of the bulk part decay slowly.
The first eigenvalue is detached from the rest of the spectrum,
and it shows an anomalous growth rate.
Eigenvalues with large magnitudes belong to eigenvectors
localized on vertices with many neighbors.
In the present context, 
{\it the low number of triangles, the high number of loops with
length above $l=4$, and the buildup of correlations are the basic
properties of the scale-free model.}

\subsection{Testing the structure of a ``real-world'' graph}
\label{ss_test}

To analyze the {\it structure
of a large sparse random graph} (correlated or not), 
here we suggest several tests,
that can be performed within ${\mathcal O}(N)$ CPU time,
use ${\mathcal O}(N)$ floating point operations, 
and can clearly differentiate
between the three ''pure'' types of random graph models treated in 
Section\,\ref{s_res}. 
Furthermore, these tests allow one to
{\it quantify the relation between any real-world graph
and the three basic types of random graphs}.

\subsubsection{Extremal eigenvalues}
\label{sss_ext}

In Section\,\ref{sss_ee} we have already mentioned that the extremal
eigenvalues contain useful information on the structure of the graph.
As the spectra of
uncorrelated random graphs (Fig.\,\ref{fig_ER})
and scale-free networks (Fig.\,\ref{fig_BA.all})
show, the principal eigenvalue of random graphs 
is often detached from the rest of the spectrum.
For these two network types,
the remaining bulk part of the spectrum 
-- i.e., the set \{$\lambda_2$, ..., $\lambda_N$\} --
converges to a symmetric distribution,
thus the quantity

\bea
R := {\lambda_1 - \lambda_2 \over \lambda_2 - \lambda_N}
\label{eq_lamb}
\eea

\noindent 
measures 
{\it the distance of the first eigenvalue from the main part of
$\rho(\lambda)$} 
normalized by the extension of the main part. 
($R$ can be connected to the chromatic 
number of the graph\cite{chromnum}.)

\unitlength10mm 
\begin{figure}
\begin{center}
\begin{picture}(9,6.2)
\put(-0.2,2.2){\psfig{width=2.5\unitlength,angle=90,file=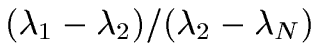}}
\put(4.6,0){$N$}
\put(0,6.2){\psfig{width=6\unitlength, angle=-90,
file=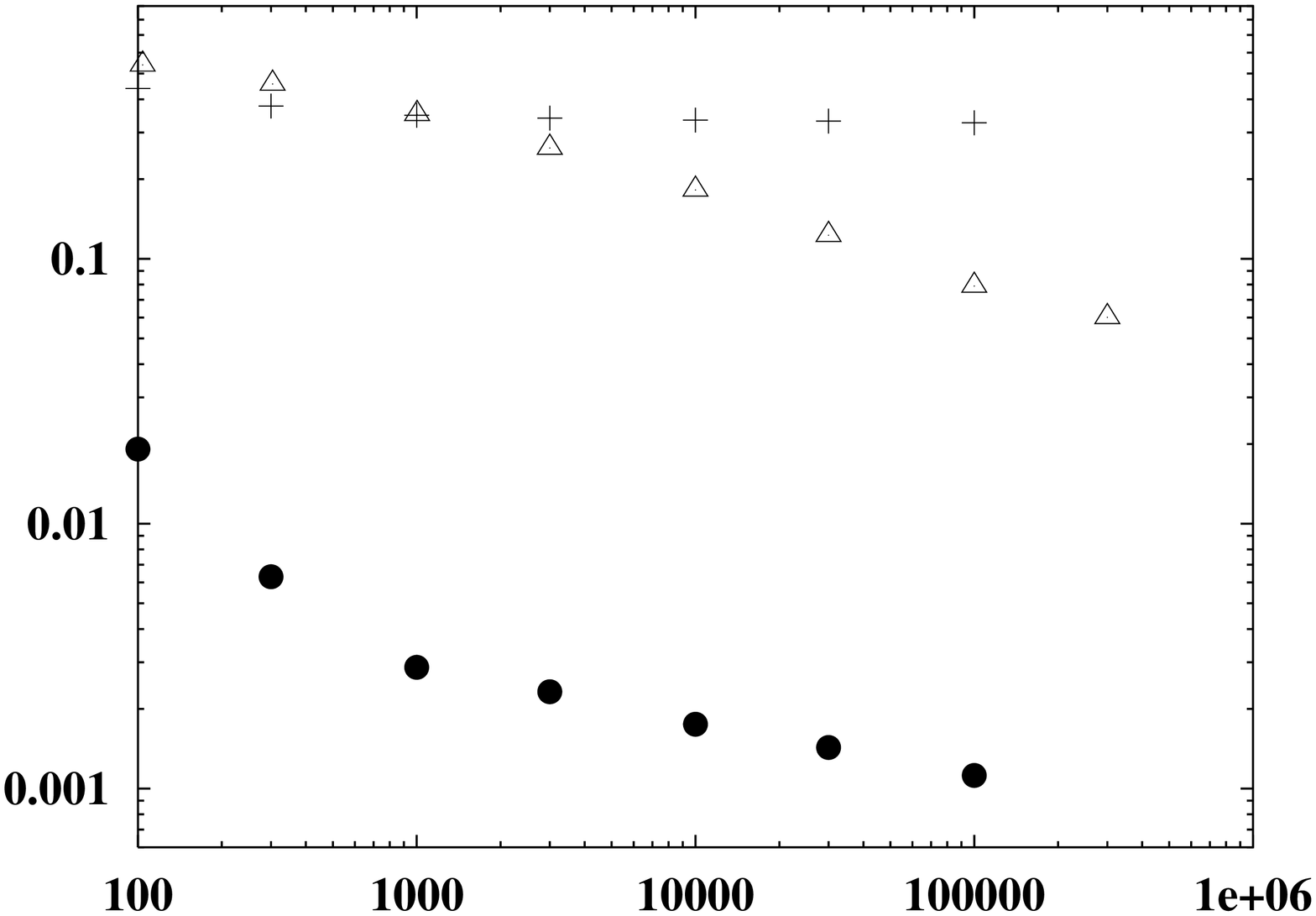}}
\end{picture}
\end{center}
\caption[]{
The ratio 
$R=(\lambda_1-\lambda_2)/(\lambda_2-\lambda_N)$ for
sparse uncorrelated random graphs ($+$), 
small-world graphs with $p_{\rm r}=0.01$ ($\bullet$) and 
scale-free networks ($\triangle$).
All graphs have an
average degree of $\langle k_i\rangle =10$, 
and at each data point,
the number of graphs used for averaging was $9$.
Observe, that for the uncorrelated random graph, 
$R$ converges to a constant (see Section\,\ref{sss_ee}),
whereas it decays rapidly for the two other types of networks, 
as $N\to\infty$. 
On the other hand, the latter two network types (small-world and
scale-free) differ significantly in their magnitudes of $R$.
\label{fig_LAMBDATEST}
}
\end{figure}

Note, that in the $N\to\infty$ limit the $\alpha=0$ sparse
uncorrelated random graph's principal eigenvalue will scale
as $\langle k_i\rangle$, whereas 
both $\lambda_2$ and $|-\lambda_N|$ will scale as 
$2\sqrt{\langle k_i\rangle}$.
Therefore, if $\langle k_i\rangle > 4$, the principal eigenvalue will
be detached from the bulk part of the spectrum and $R$ will scale
as $(\sqrt{\langle k_i\rangle} - 2 ) / 4$.
If, however, $\langle k_i\rangle \le 4$,
$\lambda_1$ will not be detached from the bulk part, 
and it will converge to $0$.

\par
The above explanation and Fig.\,\ref{fig_LAMBDATEST} show
that in the $\langle k_i\rangle > 4$
sparse uncorrelated random graph model 
and the scale-free network, 
$\lambda_1$ and the rest of the spectrum are well separated,
which gives similarly high values for $R$ in small systems. In large
systems, $R$ of the sparse 
uncorrelated random graph converges to a constant,
while $R$ in the scale-free model decays as a power law function of $N$.
The reason for this drop is the
increasing denominator on the rhs. of Eq.\,\ref{eq_lamb}:
$\lambda_2$ and $\lambda_N$ are the extremal eigenvalues in
the lower and upper long tails of $\rho(\lambda)$,
therefore, as $N$ increases, the expectation values of
$\lambda_2$ and $-\lambda_N$ grow as quickly as that of $\lambda_1$.
On the other hand, the small-world network
shows much lower values of $R$ already for small systems:
here, $\lambda_1$ is not detached from the rest of the spectrum,
which is a consequence of the almost periodical structure of the graph.

\par
On Fig.\,\ref{fig_LAMBDATEST}, graphs with the same number
of vertices and edges are compared.
For large ($N\ge 10,000$) systems, 
for sparse uncorrelated random graphs $R$ 
{\it converges to a constant}, 
whereas for scale-free graphs and
small-world networks it {\it decays as a power-law}. 
The latter two networks significantly
differ in the magnitude of $R$.
In summary, the suggested quantity $R$ 
has been shown to be appropriate for
distinguishing between the following graph structures:
(i) periodical or almost periodical (small-world), 
(ii) uncorrelated non-periodical
and (iii) strongly correlated non-periodical (scale-free).

\subsubsection{Inverse participation ratios of extremal eigenpairs}
\label{sss_iprext}

\begin{figure}
\begin{center}
\begin{picture}(9,6.2)
\put(-0.2,2.7){\psfig{width=2.5\unitlength,angle=90,file=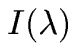}}
\put(4.5,0){$\lambda$}
\put(0,6.2){\psfig{width=6\unitlength, angle=-90, file=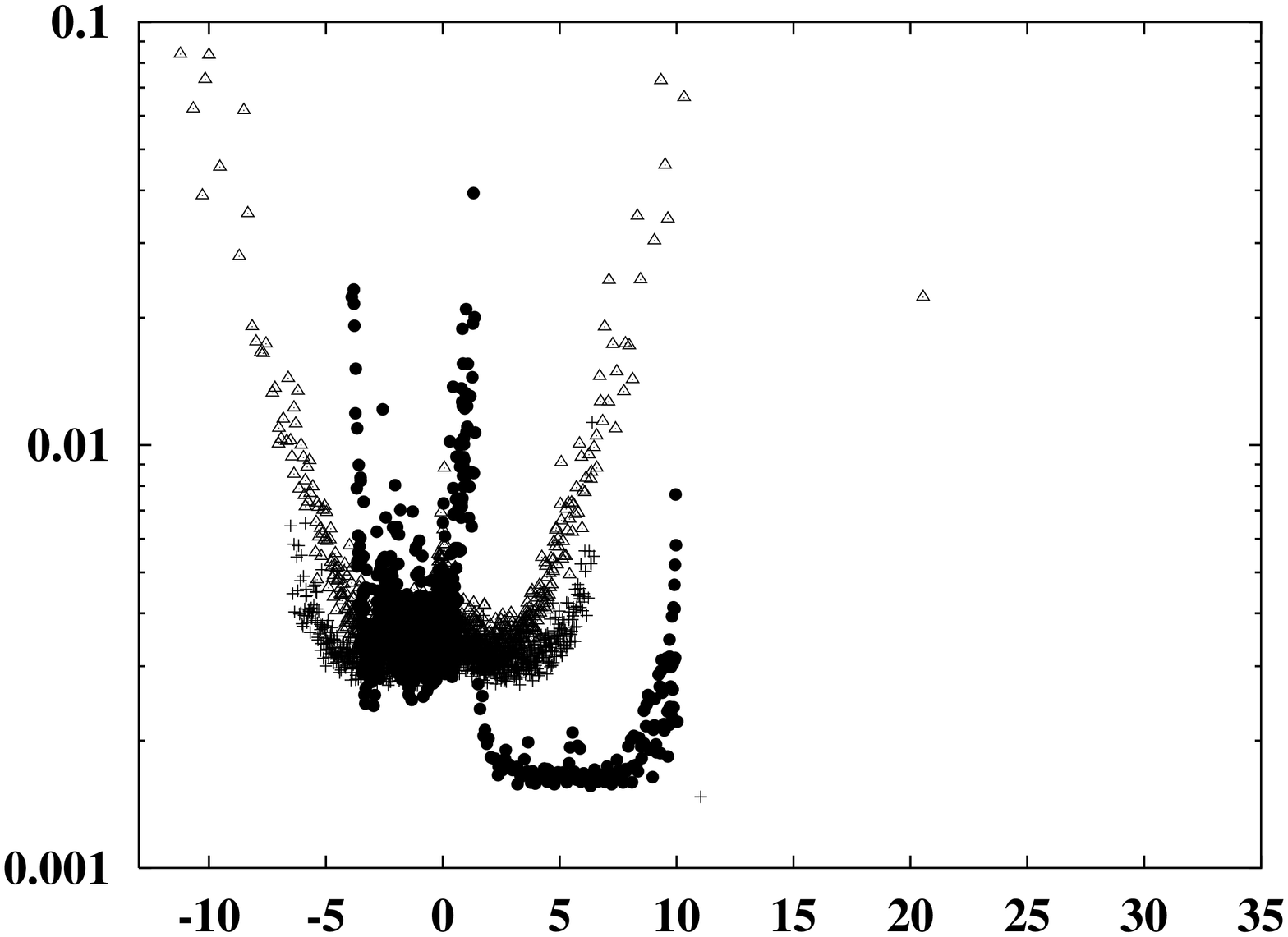}}
\put(5,3.1){\psfig{width=2.3\unitlength, angle=-90,
file=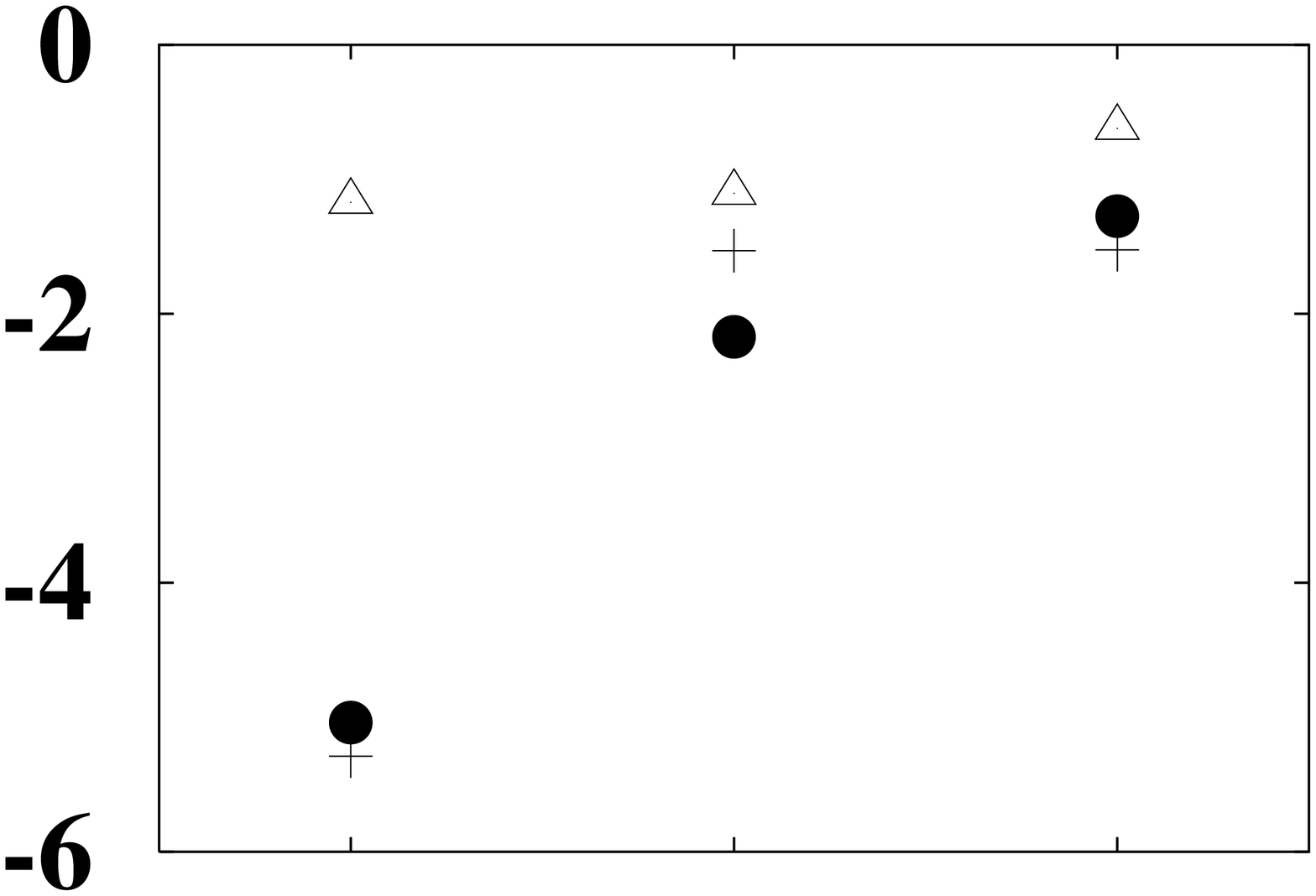}}
\put(4.8,1.5){\psfig{width=1.5\unitlength,angle=90,file=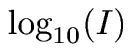}}
\put(5.9,1){\tiny ${\mathbf{\evec_1}}$}
\put(6.7,1){\tiny ${\mathbf{\evec_2}}$}
\put(7.5,1){\tiny ${\mathbf{\evec_N}}$}
\end{picture}
\end{center}
\caption[]{
{\bf Main panel:} 
Inverse participation ratios of the eigenvectors of three graphs
shown as a function of the corresponding eigenvalues: 
uncorrelated random graph ($+$), 
small-world graph with $p_{\rm r}=0.01$ ($\bullet$) 
and scale-free graph ($\triangle$). 
All three graphs have $N=1,000$ vertices, 
and the average degree of a vertex
is $\langle k_i\rangle=10$. 
Observe, that the eigenvectors of 
the sparse uncorrelated random graph and the small-world network
are usually non-localized ($I(\lambda)$ is
close to $1/N$). On the contrary, 
eigenvectors belonging to the scale-free graph's 
extremal eigenvalues are highly localized with 
$I(\lambda)$ approaching $0.1$.
Note also, that for $\lambda\approx 0$, 
the scale-free graph's $I(\lambda)$ 
has a significant ``spike'' indicating again 
the localization of eigenvectors.
{\bf Inset:} Inverse participation ratios 
of the $1$st, $2$nd and $N$th
eigenvectors of an uncorrelated random graph ($+$), 
a small-world graph with $p_{\rm r}=0.01$ ($\bullet$) 
and a scale-free graph ($\triangle$). 
For each data point,
the number of vertices was $N=300,000$ and
the number of edges was $1,500,000$.
Clearly, the principal eigenvector of the scale-free graph is
localized, while the principal eigenvector of the other two systems
(the uncorrelated models) is not. Note also, that the inverse
participation ratios of the $2$nd and $N$th eigenvectors clearly
differ in the small-world graph -- the spectrum of this graph 
has already been shown to be strongly asymmetric, -- whereas the inverse
participation ratios of $\evec_2$ and $\evec_N$ are approximately the
same in the uncorrelated random graph.
\par
Thus, with the help of the inverse participation ratios of 
$\evec_1$, $\evec_2$ and $\evec_N$, one can 
identify the three main types of random graphs used here.
\label{fig_IPR}
}
\end{figure}

Figure\,\ref{fig_IPR} shows the inverse participation
ratios of the eigenvectors 
of an uncorrelated random graph, a small-world graph with
$p_{\rm r}=0.01$ and a scale-free graph. 
Even though all three graphs have the same number of vertices 
($N=1,000$) and edges ($5,000$), 
one can observe rather specific features 
(see also the inset of Fig.\,\ref{fig_IPR}). 

\par
The {\it uncorrelated random graph's} eigenvectors show very little
difference in their level of localization, 
except for the principal
eigenvector, which is much less localized 
than the other eigenvectors; 
$I(\lambda_2)$ and $I(\lambda_N)$ are almost equal.
For the {\it small-world graph's} eigenvectors, 
$I(\lambda)$ has many different plateaus and spikes; 
the principal eigenvector is not localized,
and the $2$nd and $N$th eigenvectors
have high, but different $I(\lambda)$ values.
The eigenvectors belonging to the 
{\it scale-free graph's} largest and smallest eigenvalues
are localized on the ``largest'' vertices.
The long tails of the bulk part of $\rho(\lambda)$ 
are due to these vertices.
All three investigated eigenvectors 
($\evec_1$, $\evec_2$ and $\evec_N$)
of the scale-free graph are highly localized.
Consequently, the inverse participation ratios of the eigenvectors
$\evec_1$, $\evec_2$ and $\evec_N$ are handy for the identification
of the $3$ basic types of random graph models used.

\subsection{Structural variances}
\label{ss_var}

\subsubsection{Relative variance of the principal eigenvalue for
different types of networks: the scale-free graph and self-similarity}
\label{sss_relvar}

\begin{figure}
\begin{center}
\begin{picture}(9,6.2)
\put(-0.2,2.7){\psfig{width=2.5\unitlength,angle=90,file=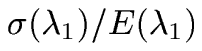}}
\put(4.5,0){$N$}
\put(0,6.2){\psfig{width=6\unitlength, angle=-90,
file=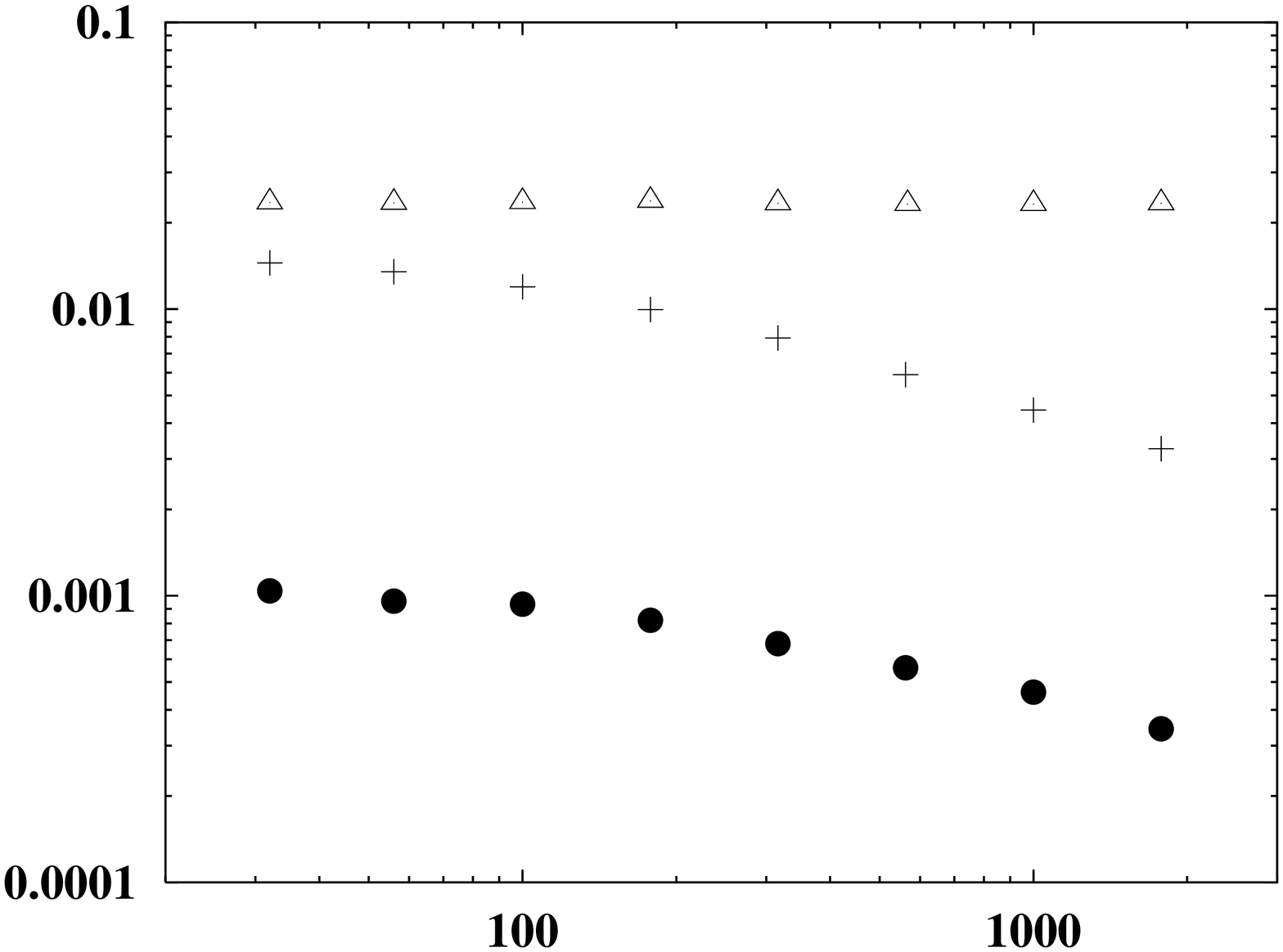}}
\end{picture}
\end{center}
\caption[]{
Size dependence of the relative variance 
of the principal eigenvalue 
-- i.e., $\sigma(\lambda_1)/E(\lambda_1)$ -- 
for sparse uncorrelated random graphs ($+$), 
small-world graphs with $p_{\rm r}=0.01$ ($\bullet$), 
and scale-free graphs ($\triangle$).
The average degree of a vertex is
$\langle k_i\rangle=10$, and $1000$ graphs were used for averaging
at every point. 
Observe, that in the uncorrelated random graph
and the small-world model
$\sigma(\lambda_1)/E(\lambda_1)$ decays
with increasing system size, however, for scale-free graphs with the
same number of edges and vertices, it remains constant.
\label{fig_sigma}
}
\end{figure}

Figure\,\ref{fig_sigma} shows the relative variance of the principal
eigenvalue -- i.e., $\sigma(\lambda_1)/E(\lambda_1)$ -- 
for the three basic random graph types.

\unitlength10mm 

For non-sparse uncorrelated random graphs 
($N\to\infty$ and $p={\rm const}$)
this quantity is known to decay at a rate which is faster than
exponential\cite{CvetRowl,McDiarmid}.
Comparing sparse graphs 
with the same number of vertices and edges, one can
see, that in the {\it sparse uncorrelated random graph}
and the {\it small-world model}
the relative variance of the principal eigenvalue 
{\it drops quickly} with growing system size.
In the {\it scale-free model}, however, the relative 
variance of the 
principal eigenvalue's distribution {\it remains constant} with
an increasing number of vertices.

\par
In fractals, fluctuations do not disappear as the size of the system
is increased, while in the scale-free graph, the relative variance of the
principal eigenvalue is independent of system size. In this sense,
the scale-free graph resembles self-similar systems.

\unitlength10mm 

\section{Conclusions}
\label{s_conc}

We have performed a detailed analysis of the 
complete spectra, eigenvalues and
the eigenvectors' inverse participation ratios
in three types of sparse random graphs: 
the sparse uncorrelated random graph, 
the small-world model
and the scale-free network.
Connecting the topological features of these graphs to algebraic
quantities, we have demonstrated that 
(i) {\it the semi-circle law is not universal}, 
not even for the uncorrelated random graph model; 
(ii) the  {\it small-world graph}
is inherently non-correlated and contains a high number of triangles;
(iii) the spectral density of the {\it scale-free graph}
is made up of three, well distinguishable parts (center, tails of
bulk, first eigenvalue), and as $N\to\infty$,
triangles become negligible and the level of correlations changes.

\par
We have presented practical tools for the
identification of the above 
mentioned basic types of random graphs
and further, for the classification of real-world graphs.
The robust eigenvector techniques and 
observations outlined in this paper
combined with previous studies
are likely to improve our understanding of 
large sparse correlated random structures.
Examples for algebraic techniques already in use
for large sparse correlated random structures 
are analyses of the Internet\cite{falou,Broder}
and search engines\cite{clever,citeseer}
and mappings\cite{Plankton,Shavitt} of the World-Wide Web.
Besides the improvement of these techniques,
the present work may turn out to be useful for analyzing
the correlation structure of the transactions
between a very high number of economical and financial
units, which has already been started in e.g., Refs.
\cite{Laloux-Bouchaud,Plerou-Stanley,Mantegna}.
Lastly, we hope to have provided quantitative 
tools for the classification of further ``real-world'' networks, 
e.g., social and biological networks.

\vskip1truecm
\centerline{\bf ACKNOWLEDGEMENTS}
\vskip1truecm

We thank D. Petz, G. Stoyan, G. Tusn\'ady, K. Wu and B. Kahng for helpful
discussions and comments.
This research was partially supported by a HNSF Grant No. OTKA T033104 and
NSF PHY-9988674.

\par
After submitting our manuscript we were made aware of a
manuscript of K.-I. Goh, B. Kahng and D. Kim~\cite{Kahng}, 
which also investigates the eigenvalues of random graph models.

\section{Appendices}
\subsection{The spectrum of a small-world graph for 
$p_{\rm r}=0$ rewiring probability}
\label{app_SW}

\subsubsection{Derivation of the spectral density}
\label{sss_der}

\begin{figure}
\begin{center}
\begin{picture}(9,4.6)
\put(1,6.2){\psfig{width=7\unitlength, angle=-90, file=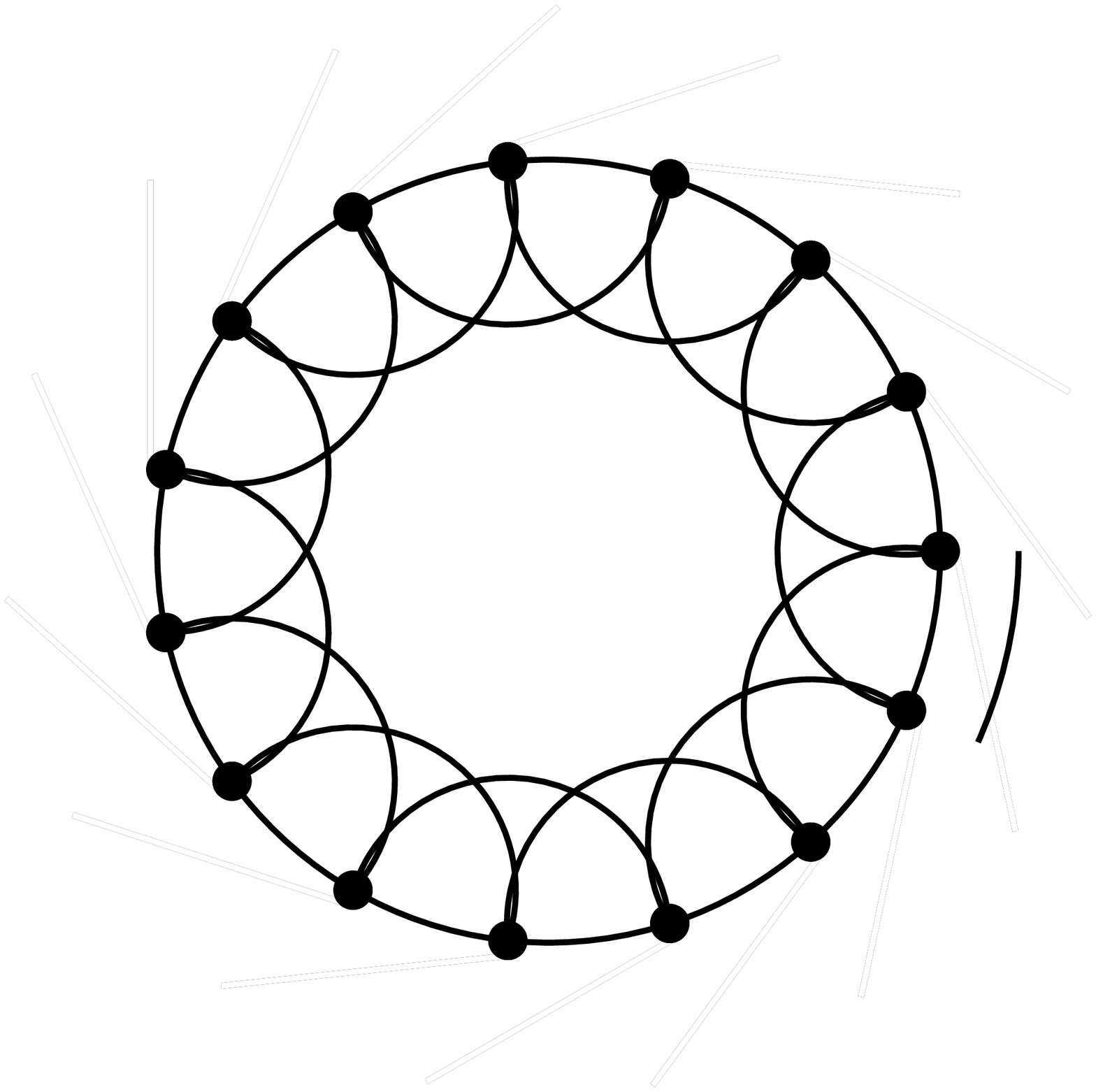}}
\put(4.5,0.6){\vector(1,0){0.01}}
\put(4,0.2){$P^{(1)}$}
\end{picture}
\end{center}
\caption[]{
The regular ring graph obtained from the small-world model in the 
$p_{\rm r}=0$ case: rotations ($P^{(n)}$ for every $n=0,1,...,N-1$) 
are symmetry operations of the graph. The  $P^{(n)}$ operators (there are
$N$ of them) can be used to create a full orthogonal basis of the
adjacency matrix, $A$:
taking any $P^{(n)}$, it commutes with $A$, therefore  
they have a common full orthogonal system of eigenvectors.
(For a clear illustration of symmetries,
this figure shows a graph with only $N=15$ vertices and $k=4$
connections per vertex.)
\label{fig_SW.symmetries}
}
\end{figure}

If the rewiring probability of a small-world graph is 
$p_{\rm r}=0$, then the
graph is regular, each vertex is connected to its $k$ nearest
neighbors and the eigenvalues can be computed using the graph's
symmetry operations. 
Rotational symmetry operations can be easily recognized, 
if the vertices of the graph are drawn 
along the perimeter of a circle (see Fig.~\ref{fig_SW.symmetries}):
let $P^{(n)}$ ($n=0,1,...,N-1$)
denote the symmetry operation that
rotates the graph by $n$ vertices in the anticlockwise direction.
Being a symmetry operation, each $P^{(n)}$ commutes with the adjacency
matrix, $A$, and they have a common full orthogonal system of
eigenvectors.

\par
Now, we will create a full orthogonal basis of $A$.
(We will treat only the
case when $N$ is an even number; odd  $N$'s
can be treated similarly.) It is known, that the eigenvalues of $A$
are real, however, to simplify calculations, we will use complex
numbers first. 
The eigenvectors of every $P^{(n)}$ are 
$\evec_1,\evec_2,...,\evec_N$:

\bea
(e_l)_j = \exp\bigg(2\pi i\, \frac{jl}{N}\bigg) \, ,
\eea

where $l=0,2,...,N-1$ and $i=\sqrt{-1}$. 
The eigenvalue of $P^{(n)}$ on $\evec_l$ is 

\bea
s_l^{(n)}=\exp\bigg(2\pi i\,\frac{nl}{N}\bigg) \, .
\eea

\unitlength10mm 

By adding these values pairwise, 
one can obtain the $N$ eigenvalues of the graph:

\bea
\lambda_l= 2\sum_{j=1}^{k/2} \
\cos\bigg(2\pi\,\frac{jl}{N}\bigg) \, .
\label{eq_appSW.lambda1.1}
\eea

In the previous exponential form 
the rhs. is a summation for a geometrical series,
therefore

\bea
\lambda_l = \
{\sin  [(k+1) l \pi / N] \over \sin(l\pi/N) } - 1 \, .
\eea

In the $N\to\infty$ limit, this converges to

\bea
\lambda(x)={\sin[(k+1)x]\over\sin(x)} -1 \, ,
\label{eq_lamx}
\eea

where $x$ is evenly distributed in the 
interval $[ 0,\pi [$.

\subsubsection{Singularities of the spectral density}
\label{sss_sing}

The spectral density is singular in $\lambda =\lambda(x)$, 
if and only if ${d\lambda\over dx} (x)=0$, which is equivalent to 

\bea
(k+1)\tan (x) = \tan [(k+1)x] \, .
\label{eq_tan}
\eea

Since $k$ is an even number,
both this equation and Eq.\,(\ref{eq_lamx}) are invariant 
under the transformation $x\mapsto\pi-x$, therefore only the 
$x\epsilon [ 0,\pi/2 ]$ solutions will give different
$\lambda$ values.
If $k=10$ (see Fig.\,\ref{fig_SW.p}), 
Eq.\,(\ref{eq_tan}) has $k/2+1=6$ solutions in $[ 0,\pi/2 ]$,
which are
$x$ $=$ $0$, $0.410$, $0.704$, $0.994$, $1.28$ and $\pi/2$.
Therefore -- according to Eq.\,(\ref{eq_lamx}) --
in the $N\to\infty$ limit, the spectral density will be singular in
the following points:

\bea
\lambda_i = -3.46,\,-2.19,\,-2,\,0.043,\,0.536,\,{\rm and}\,k=10\,. 
\eea

%

\subsection{Crossover in the growth rate of the scale-free graph's
principal eigenvalue}
\label{app_crossover}

The largest eigenvalue is influenced only by the longest row vector if
and only if the two longest row vectors are almost orthogonal:

\bea
\vec{v}_1 \vec{v}_2 \ll |\vec{v}_1| |\vec{v}_2| \, .
\label{eq_vv}
\eea

For $m>1$, the lhs. of (\ref{eq_vv})
is the number of simultaneous 
$1$'s in the two longest row vectors, 
and the rhs. can be approximated with 
$|\vec{v}_1|^2=k_1$, the largest degree of the graph.
It is known\cite{ScaleFree_PA}, that for large $j$ ($j>i$), 
the $j$th vertex will be connected to 
vertex $i$ with probability $P_{ij}=m/(2\sqrt{ij})$.
Thus, we can write (\ref{eq_vv}) in the following forms:

\bea
\sum_{t=1}^N P_{1t}^2 \ll \sum_{t=1}^N P_{1t} \, ,
\eea

or

\bea
{\sqrt{N_{\rm c}}\over\ln{N_{\rm c}}} \gg {m\over 4}
\eea

where $N_{\rm c}$ is the critical system size.


\end{multicols}


\begin{references}
\bibitem{def_regular}
The number of edges meeting at one vertex of a graph
is called the degree of that 
vertex, and a graph is called regular, if every vertex has the same
degree.
\bibitem{alpha=1}
If $n_{\rm s,e}$ denotes the number of isolated clusters with $s$
vertices and $e$ edges, then the number of isolated clusters of size
$s$ can be written as $n_{\rm s}=\sum_{e=s-1}^{s(s-1)/2} n_{\rm s,e}$.
\par
The $e=s-1$ case corresponds to stars of size $s$, and 
$n_{\rm s,s-1}$ can be given as
$\big( {N\atop s} \big) s p^{s-1} (1-p)^{ s(N-s)+{s(s-1)\over 2}-(s-1)}$.
If $N\to\infty$ and $pN\to c$, 
then for any finite $s$ this converges to $N f(s)$, where
$f(s)=(2\pi)^{-1/2} s^{-s+1/2} c^{s-1} \exp(s-sc)$, 
which is an exponentially decreasing function of $s$, if $c>1$.
\par
Similarly to this case, in the $N\to\infty$ and $pN\to c$ limit,
for all values of $e$, $n_{\rm s,e}$ is
an exponentially decaying function of $s$, if $c>1$.
Since the number of possible $e$ values for a fixed $s$ is a
polynomial function of $s$, $n_{\rm s}$ will also decay exponentially,
if $c>1$, $N\to\infty$ and $pN\to c$.
\bibitem{delta}
In a graph with $N$ vertices,
the contribution of one isolated vertex to the spectral density is 
$N^{-1}\delta(\lambda)$, 
and the contribution of an isolated cluster with two vertices is 
$N^{-1}[\delta(\lambda+1)+\delta(\lambda-1)]$.
An isolated cluster with three vertices and two edges will give 
$N^{-1}[\delta(\lambda+\sqrt{2})+\delta(\lambda)+\delta(\lambda-\sqrt{2})$,
and one with three vertices and three edges adds
$N^{-1}[2\delta(\lambda+1)+\delta(\lambda-2)]$ to the spectral density.
\end{references}
\end{document}